%% file: main.tex
\newif\iffull
\fulltrue

\documentclass[sigconf, nonacm]{acmart}
\usepackage[table]{xcolor}
\input{macro}
\newcommand\vldbdoi{XX.XX/XXX.XX}
\newcommand\vldbpages{XXX-XXX}
\newcommand\vldbvolume{17}
\newcommand\vldbissue{1}
\newcommand\vldbyear{2024}
\newcommand\vldbauthors{\authors}
 
\newcommand\vldbavailabilityurl{https://github.com/ParAlg/ParClusterers}
\newcommand\vldbpagestyle{plain}

\begin{document}
\title{The ParClusterers Benchmark Suite (PCBS): \\
A Fine-Grained Analysis of Scalable Graph Clustering \\
{[}Experiment, Analysis \& Benchmark{]}}

\author{Shangdi Yu}
\affiliation{%
  \institution{MIT}
}
\email{shangdiy@mit.edu}

\author{Jessica Shi}
\affiliation{%
  \institution{MIT}
}
\email{jessicashi@gmail.com}

\author{Jamison Meindl}
\affiliation{%
  \institution{MIT}
}
\email{jmeindl@mit.edu}

\author{David Eisenstat}
\affiliation{%
  \institution{Google}
}
\email{eisen@google.com}

\author{Xiaoen Ju}
\affiliation{%
  \institution{Google}
}
\email{xiaoen@google.com}

\author{Sasan Tavakkol}
\affiliation{%
  \institution{Google}
}
\email{tavakkol@google.com}

\author{Laxman Dhulipala}
\authornote{These authors are ordered alphabetically.}
\affiliation{%
  \institution{UMD}
}
\email{laxman@umd.edu}

\author{Jakub {\L}{\k{a}}cki}
\authornotemark[1]
\affiliation{%
  \institution{Google}
}
\email{jlacki@google.com}

\author{Vahab Mirrokni}
\authornotemark[1]
\affiliation{%
  \institution{Google}
}
\email{mirrokni@google.com}

\author{Julian Shun}
\authornotemark[1]
\affiliation{%
  \institution{MIT}
}
\email{jshun@mit.edu}

\begin{abstract}
We introduce the ParClusterers Benchmark Suite (PCBS)---a collection of highly scalable parallel graph clustering algorithms and benchmarking tools that streamline comparing different graph clustering algorithms and implementations.
The benchmark includes clustering algorithms that target a wide range of modern clustering use cases, including community detection, classification, and dense subgraph mining.
The benchmark toolkit makes it easy to run and evaluate multiple instances of different clustering algorithms, which can be useful for fine-tuning the performance of clustering on a given task, and for comparing different clustering algorithms based on different metrics of interest, including clustering quality and running time.

Using PCBS, we evaluate a broad collection of real-world graph clustering datasets.
\revised{Somewhat surprisingly, we find that the best quality results are obtained by algorithms that not included in many popular graph clustering toolkits.}
The PCBS provides a standardized way to evaluate and judge the quality-performance tradeoffs of the active research area of scalable graph clustering algorithms. We believe it will help enable fair, accurate, and nuanced evaluation of graph clustering algorithms in the future.

\end{abstract}

\maketitle

\pagestyle{\vldbpagestyle}
\begingroup\small\noindent\raggedright\textbf{PVLDB Reference Format:}\\
\vldbauthors. 
The ParClusterers Benchmark Suite (PCBS): 
A Fine-Grained Analysis of Scalable Graph Clustering 
{[}Experiment, Analysis \& Benchmark{]}. PVLDB, \vldbvolume(\vldbissue): \vldbpages, \vldbyear.\\
\href{https://doi.org/\vldbdoi}{doi:\vldbdoi}
\endgroup
\begingroup
\renewcommand\thefootnote{}\footnote{\noindent
This work is licensed under the Creative Commons BY-NC-ND 4.0 International License. Visit \url{https://creativecommons.org/licenses/by-nc-nd/4.0/} to view a copy of this license. For any use beyond those covered by this license, obtain permission by emailing \href{mailto:info@vldb.org}{info@vldb.org}. Copyright is held by the owner/author(s). Publication rights licensed to the VLDB Endowment. \\
\raggedright Proceedings of the VLDB Endowment, Vol. \vldbvolume, No. \vldbissue\ %
ISSN 2150-8097. \\
\href{https://doi.org/\vldbdoi}{doi:\vldbdoi} \\
}\addtocounter{footnote}{-1}\endgroup

\ifdefempty{\vldbavailabilityurl}{}{
\vspace{.3cm}
\begingroup\small\noindent\raggedright\textbf{PVLDB Artifact Availability:}\\
The source code, data, and/or other artifacts have been made available at \url{\vldbavailabilityurl}.
\endgroup
}

\input{intro}
\input{api}

\input{exp}

\input{primer}

\section{Conclusion}

\revised{
The ParClusterers Benchmark Suite (\parc{}) provides a comprehensive platform for evaluating scalable parallel graph clustering algorithms. Our experiments yielded three key insights: (1) many of \parc{}'s clustering implementations are significantly faster than those of existing graph libraries and databases, while obtaining very good, and in many cases the best, clustering quality; (2) 
in the 4 tasks we studied, the best overall quality (AUC) is obtained by either correlation clustering or ParHAC. We note that these methods are not included in many popular graph algorithms packages (see \Cref{tab:all_platforms}), and we hope that our positive results encourage more research on their quality and practical applicability.
(3) PCBS's modularity-based methods are competitive or superior in quality and speed compared to other implementations. These findings highlight \parc{}'s value in enabling thorough, standardized comparisons of clustering algorithms.

\parc{}'s modular design allows users to easily extend it with new algorithms, datasets, or metrics, fostering innovation in graph clustering research. Researchers and practitioners can integrate their own custom clustering algorithms and datasets into \parc{}, enabling direct comparisons with state-of-the-art methods on both standard and domain-specific graphs.

} 

Besides the clustering applications explored in this work, there are also many other interesting future directions. One idea is to study how weighted graph clustering algorithms perform when derived weights are added to the unweighted graph (for example, weights derived from node degrees as in \cite{dhulipala2023terahac}). 
Moreover, we observe a big gap between the clustering time of the slower algorithms (e.g., correlation clustering and ParHAC) and the fastest (e.g., LDD and connectivity).  It would be interesting to study if we can close the gap by developing much faster clustering algorithms with close-to-best clustering quality that obtain the best of both worlds.

\clearpage

\bibliographystyle{ACM-Reference-Format}
\bibliography{sample}

\iffull
\appendix
\input{appendix-scalability}

\input{appendix-k}

\input{appendix-datasets}

\input{appendix-exp}

\fi

\end{document}

%% file: macro.tex
\usepackage[utf8]{inputenc}

\usepackage{pgfplots}

\usepackage{amsmath, amsfonts}
\usepackage{cleveref}

\usepackage{xy}
\usepackage{tikz}
\usepackage{mathrsfs}
\usepackage{setspace,dsfont,graphicx,makecell,appendix,multirow,float}
\pgfplotsset{compat=1.17}

\usepackage{bbm}
\usepackage{graphicx}

\usepackage{tabularx}
\usepackage{dblfloatfix}
\usepackage{xspace}
\usepackage{algorithm}
\usepackage[noend]{algpseudocode}
\usepackage{subcaption,graphicx,xurl,multicol,mathtools}
\usepackage{paralist}
\usepackage{color}
\usepackage{arydshln}
\usepackage[most]{tcolorbox}
\usepackage{stmaryrd}
\usepackage{booktabs}
\usepackage{siunitx}
\usepackage{bm}
\usepackage{enumitem}
\setlist{nosep}
\graphicspath{ {./figures/} }
\usepackage{balance}
\usepackage[font=small,labelfont=bf]{caption}

\newcommand{\slfrac}[2]{\left.#1\middle/#2\right.}

\newcommand{\whp}[1]{w.h.p.}

\algblock{ParFor}{EndParFor}
\algnewcommand\algorithmicparfor{\textbf{parfor}}
\algnewcommand\algorithmicpardo{\textbf{do}}
\algnewcommand\algorithmicendparfor{}%
\algrenewtext{ParFor}[1]{\algorithmicparfor\ #1\ \algorithmicpardo}
\algrenewtext{EndParFor}{\algorithmicendparfor}
\algtext*{EndParFor}

\algdef{SE}[DOWHILE]{Do}{doWhile}{\algorithmicdo}[1]{\algorithmicwhile\ #1}%

\newcommand{\myparagraph}[1]{\vspace{1pt}\noindent {\bf #1.}}
\usepackage{comment}

\newcommand{\parc}{PCBS}
\newcommand{\knn}{$k$\text{-nearest} \text{neighbor}\xspace}

\newcommand{\savespace}[1]{}

\newcommand{\datasetname}[1]{\texttt{#1}}

\usepackage[compact]{titlesec}
\titlespacing*{\section}{0pt}{*1}{*1}
\titlespacing*{\subsection}{0pt}{*1}{*1}
\titlespacing*{\subsubsection}{0pt}{*1}{*1}

\newcommand{\revised}[1]{#1}

%% file: intro.tex
\section{Introduction}
Clustering is a critical tool in almost any scientific field that
involves classifying and organizing data today.
Examples of fields leveraging clustering range from
computational biology and phylogenetics to complex network analysis,
machine learning, and astrophysics~\cite{irbook, shalita2016social, Camerra2014,patwary2015bd}.
Clustering has proven particularly useful in fields transformed by AI
and machine learning because of its utility in understanding and
leveraging high-dimensional vector representations (embeddings) of
data~\cite{monath2021scalable, bateni2017affinity, faiss, dhulipala2021hierarchical, dhulipala2022hierarchical}.

\revised{In this paper, we are interested in carefully characterizing the behavior (e.g., measuring quality, running time, and scalability) of parallel clustering algorithms for shared-memory multi-core machines that are scalable in the size of the dataset and the number of threads.
Our specific focus is on {\em graph clustering}, which is a versatile and scalable clustering approach that can be used with different input types.}
On one hand, graph clustering is a natural approach whenever the input is a graph (e.g., friendships, interactions, etc.).
On the other hand, graph clustering can also be applied in the other popular scenario, when the input is a collection of points in a metric space (e.g., embeddings).
In this case, one can obtain a graph by computing a {\em weighted similarity graph}, where continuous or complete
phenomena can be cast into sparse similarity graphs, e.g., by keeping
only edges between nearby points or only the most significant entries
of a similarity matrix.

Despite substantial prior works that study the quality (e.g., precision
and recall) and scalability of individual graph clustering methods~\cite{dhulipala2023terahac, dhulipala2021hierarchical, dhulipala2022hierarchical, shi2021scalable, staudt2016networkit, tsourakakis2017scalable, tseng2021parallel}, \revised{no
prior works have systematically compared a large collection of different graph clustering
methods (and their corresponding implementations) to understand how different methods compare against each other
under different metrics.}
For example, celebrated and widely-utilized graph clustering
algorithms, such as modularity clustering are well understood to be
highly effective in community detection tasks on unweighted natural
graphs, but little is known about their performance for
clustering on vector embedding clustering tasks.

This paper addresses this gap by performing a systematic comparison of a large and diverse set of graph clustering methods.
Our evaluation includes methods tailored to both weighted and unweighted
graphs and incorporates a diverse set of natural graphs and similarity graphs derived from point sets.
\revised{
We focus on undirected graphs, as 
converting directed graphs to undirected graphs is a common practice in many graph tasks, such as community detection (see, e.g.,~\cite{fortunato2010community}).}
We stratify our evaluation based on \revised{four} unsupervised
clustering tasks that are commonly found in the literature and in
practice---(1) community detection, (2)
vector embedding clustering, (3) dense subgraph partitioning, \revised{ and (4) high resolution clustering.}
Due to insisting on scalability, we focus our evaluation on the most
scalable parallel graph clustering methods currently available in the
literature.

To make our evaluation easily reusable and extensible by future
researchers, we designed a graph clustering benchmark called the {\bf ParClusterers Benchmark Suite (\parc)}.
\parc{} enables users to accurately measure the scalability and
accuracy of different shared-memory parallel graph clustering algorithms.
In addition to providing a simple and easy to use benchmarking
platform, we have also incorporated eleven parallel graph
clustering methods into \parc{}.
The algorithms include algorithms from our recent prior work, as well as several new implementations.
In addition to classic graph clustering methods such as
modularity-based clustering~\cite{shi2021scalable}, structural clustering~\cite{Tseng2021},
and label propagation methods~\cite{raghavan2007near}, we include recently developed
hierarchical agglomerative graph clustering methods~\cite{dhulipala2022hierarchical} and
connectivity-based methods such as $k$-core and low-diameter
decomposition~\cite{gbbs}.
\revised{Finally, unlike much of the existing work on graph clustering, which
typically focuses on optimizing a specific graph clustering metric (e.g., modularity
or conductance) that a clustering method is usually designed to
optimize, \parc{} supports evaluating any clustering algorithm using a very broad set of metrics,
which helps us understand {\em what} different clustering algorithms
are able to optimize for on real-world datasets, and help inform users
of the best clustering algorithm for a given metric.}
Besides \parc{}'s clustering implementations, \parc{} also supports running many clustering implementations in other graph clustering frameworks and systems such as NetworKit~\cite{staudt2016networkit}, Neo4j~\cite{neo4j}, and TigerGraph~\cite{tigergraph}.
\revised{\parc{} can also be easily extended to include new datasets, algorithms, and parameter search methods.}

The datasets we study include both widely-used graph
datasets from the SNAP repository, as well as several new graph
clustering datasets that we have generated from spatial and embedding datasets using
a simple nearest-neighbor-based graph building process, and which we
will open-source as part of this work.
\revised{We also contribute a new graph dataset for clustering, which represent similarities between 1.2M short texts. As far as we know, this is the first large-scale graph clustering dataset that provides a large number of ground-truth clusters.}
Our datasets cover a wide range of scales and clustering tasks, including community detection, vector embedding clustering, and dense subgraph partitioning.

\myparagraph{Key Contributions} The key contributions of our work include:
\begin{itemize}[leftmargin=*]
\item A comprehensive library that implements eleven state-of-the-art scalable graph clustering algorithms, providing a unified codebase for researchers and practitioners.

\item A benchmarking {\em toolkit} that facilitates the systematic evaluation of graph clustering algorithms across diverse datasets, parameter settings, and experimental configurations, enabling rigorous and comprehensive comparative analyses. 

\item \revised{A new large graph clustering dataset containing many ground-truth clusters.}

\item  The first extensive evaluation of parallel graph clustering algorithms, encompassing their runtime performance, clustering quality, and the trade-off between these two critical dimensions. We also compare our library against other existing libraries and graph databases.

\end{itemize}

\vspace{5pt}
\myparagraph{Key Results}
Some of our key takeaways and findings of our study of graph clustering algorithms include:
\begin{itemize}[leftmargin=*]
\item Our clustering implementations in \parc{} are very fast compared to other clustering implementation in state-of-the-art graph libraries and databases. 
While graph databases, such as Neo4j~\cite{neo4j} and TigerGraph~\cite{tigergraph}, provide a richer functionality, on different graph clustering implementations, they are slower. For example, on the LiveJournal graph from SNAP~\cite{leskovec2016snap}, \parc{} is on average 32.5x faster than Neo4j and 303x faster than TigerGraph.
Compared with state-of-the-art parallel graph library NetworKit~\cite{staudt2016networkit}, \parc{} is on average 4.54x faster. We compute the average using the geometric mean.
\item \revised{Correlation clustering~\cite{shi2021scalable} obtains the highest quality on three out of four tasks. ParHAC~\cite{dhulipala2022hierarchical} obtains the best quality on the fourth task. 
We consider this finding surprising, given that these two methods are not included in many  popular graph clustering frameworks.
The best performing method commonly found in existing graph clustering packages is modularity clustering.
However, we observe that on a vast majority of datasets, correlation clustering obtains strictly better quality.
}

\item Parallel affinity clustering obtains high quality on the vector embedding clustering task and is consistently faster than correlation clustering and ParHAC on large graphs.

\end{itemize}

\revised{Our code and the full version of our paper can be found at \url{https://github.com/ParAlg/ParClusterers}.}

%% file: api.tex
\section{ParClusterers Benchmark Suite}
\begin{figure}
    \centering
    \includegraphics[width=0.8\columnwidth]{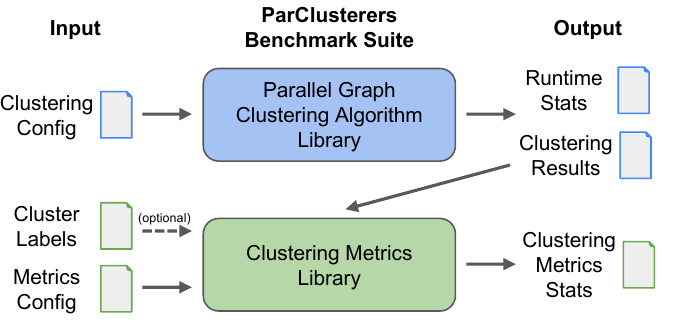}
    \vspace{-1.5em}
    \caption{Overview of our \parc{} library.}
    \label{fig:overview}
\end{figure}

In this section, we introduce our \parc{} benchmark suite, which includes a large number of scalable parallel graph clustering algorithms, parallel implementations of clustering evaluation metrics, efficient graph I/O routines, and a convenient benchmark interface. An overview of the library is shown in \Cref{fig:overview}.

\subsection{Graph Input}
\parc{} supports both \textbf{weighted and unweighted simple undirected graphs} in edge list format~\cite{leskovec2016snap} and compressed sparse row format~\cite{gbbs, gustavson1972some}. A simple graph is a graph with no self-loops and no parallel edges.
The graph clustering algorithms in \parc{} that use edge weights assume edge weights are nonnegative \textbf{similarities}, i.e., a higher weight of an edge $(u, v)$ means that $u$ and $v$ are more similar. This allows for a natural way of defining how a missing edge affects the clustering: in most of our algorithms a missing edge is equivalent to an edge of similarity $0$. 

\subsection{Datasets}
In this work, we benchmark on both unweighted and weighted graphs. For unweighted graphs, the SNAP~\cite{leskovec2016snap} collection of datasets is a popular choice for benchmarking community detection tasks.
To the best of the authors' knowledge, there currently exists no real-world {\em weighted} graph dataset with ground truth for evaluating weighted graph clustering algorithms. 
In this work, we present a suite of weighted graph clustering datasets that are the $k$-nearest neighbor graphs of real-world vector datasets with ground truth clustering labels, \revised{including a new dataset that we created that contains a large number of ground truth clusters.}
We describe more details of constructing these datasets in \Cref{sec:exp}. 
We believe that these datasets will allow researchers to thoroughly benchmark the performance of weighted graph clustering algorithms.

\subsection{Parallel Graph Clustering Algorithms}\label{sec:algorithms}
Our library includes eleven scalable parallel graph clustering algorithms, which are described below. We focus on algorithms that are scalable, widely used, and have high precision (i.e., most vertices in a retrieved cluster are from the same ground truth cluster).
We discuss more about our algorithm selection in \Cref{sec:scope}. We select a mix of algorithms that are popular in the literature and algorithms that we found to have high quality in our experience. 
The implementations of TECTONIC, label propagation, and speaker-listener label propagation (SLPA) are new. Other implementations are taken from previous work and integrated into \parc{}.
All implementations, except for SLPA, produce non-overlapping clusters.

We categorize the algorithms into \textit{weighted graph algorithms}, which take into account the similarities, and \textit{unweighted graph algorithms}, which only use the topology of the graph.

\subsubsection{Weighted Graph Algorithms} 
\myparagraph{Affinity Clustering~\cite{bateni2017affinity, dhulipala2021hierarchical}}
Affinity clustering is a hierarchical clustering algorithm based on Boruvka's minimum spanning forest algorithm. \parc{} includes the shared-memory parallel affinity clustering from \citet{dhulipala2021hierarchical}, which is adapted from the original MapReduce affinity clustering algorithm~\cite{bateni2017affinity}.  
In each step of the algorithm, every vertex
picks its best edge (i.e., that of the maximum similarity) on each round, and the connected components spanned by the selected edges define the clusters.
The algorithm can produce a clustering hierarchy by contracting the clusters and running the algorithm recursively.
The edge weights between contracted vertices can be computed using different linkage functions, such as single, maximum, or average linkage. 
In our experiments we use average linkage.

Users can control the resolution of the clustering by picking an edge weight threshold and only considering edges whose weight is at least the threshold in each step.
Recent work by \citet{monath2021scalable} showed that the quality of affinity clustering can be improved if the threshold decays geometrically at each step.
We use this technique, also known as the SCC algorithm~\cite{monath2021scalable}, in our experiments.

\myparagraph{Correlation and Modularity Clustering~\cite{shi2021scalable}}
Our library includes a shared-memory parallel framework for optimizing the LambdaCC objective~\cite{veldt2018correlation}.
This objective generalizes both modularity~\cite{newman2004finding} and correlation clustering~\cite{bansal2004correlation} and is defined as follows.

Let $k_v$ be the vertex weight of $v$ and $\lambda$ be the resolution parameter. Let $V$ be the set of all vertices, and $w_{uv}$ be the weight (similarity) of edge $(u,v)$.
Then, $\texttt{LambdaCC}(x) = \allowbreak \sum_{(i, j) \in V \times V} \allowbreak w'_{ij} \cdot \allowbreak x_{ij}$, where $x_{ij}$ is a Boolean indicator, which is equal to $1$ if and only if $i$ and $j$ belong to the same cluster. 
Here, $w'_{uv} = 0$ if $u = v$, $w'_{uv} = w_{uv} - \lambda k_u k_v$ if $(u, v) \in E$, and $w'_{uv} =  - \lambda k_u k_v$ otherwise. By default our framework uses $k_u = 1$ for any $u \in V$.

As shown in~\cite{shi2021scalable}, the special case of the modularity objective can be obtained by appropriately defining vertex weights $k$ and setting $\lambda$. %
Specifically, assume we set the vertex weight $k_v$ to be equal to the weighted degree of $v$ (i.e., the sum of its incident edge weight) and set the resolution $\lambda = \slfrac{1}{(2m)}$.
Then, optimizing LambdaCC objective is equivalent to optimizing the modularity objective~\cite{GiNe02}, as the two objectives are monotone in each other.
Furthermore, if we set the resolution $\lambda = \slfrac{\gamma}{(2m)}$, the LambdaCC objective becomes monotone in the generalized modularity objective of Reichardt and Bornholdt \cite{ReBo06} with a fixed scaling parameter $\gamma \in (0, 1)$.

Our framework optimizes the LambdaCC objective using a parallel Louvain-style algorithm~\cite{louvain, shi2021scalable}.

\myparagraph{Approximate Parallel Hierarchical Agglomerative Clustering (ParHAC)~\cite{dhulipala2022hierarchical}} \revised{Given $n$ vertices, a sequential version of the HAC algorithm starts by forming a separate cluster for each vertex,
and proceeds in $n-1$ steps. }
Each step merges the two most similar clusters, i.e., replaces them by their union. 
The similarity of two clusters is the total weight of edges between the clusters divided by the product of their sizes.

The output of HAC is a binary tree called a dendrogram, which describes the merges performed by the algorithm.
A flat clustering can be obtained from the dendrogram by cutting it at a given level. The threshold for cutting the dendrogram controls the clustering resolution. 
Because the output of HAC is a dendrogram, one can easily  postprocess the dendrogram to obtain a flat clustering of any given resolution.
The running time of this postprocessing is negligible compared to the clustering time.
\parc{} includes a shared-memory parallel implementation of \emph{approximate} HAC called ParHAC~\cite{dhulipala2022hierarchical}.
A HAC algorithm is called $1+\epsilon$ approximate, when each merged pair of clusters has similarity at least $W_{\max}/(1 + \epsilon)$, where $W_{\max}$ is the largest similarity between any two clusters.

\myparagraph{Connected Components}~\cite{gbbs, jayanti2016randomized}
Given a threshold parameter $\tau$, the clusters are the connected components with edge similarity $<\tau$ removed. Our implementation uses the state-of-the-art implementation of concurrent union-find capable of processing several billions of edges per second from the ConnectIt library~\cite{jayanti2016randomized, connectit} to find the connected components.

\subsubsection{Unweighted Graph Algorithms}
\myparagraph{Low-Diameter Decomposition (LDD)~\cite{gbbs, miller2013parallel}}
LDD partitions vertices of an $n$-vertex, $m$-edge unweighted graph into clusters, where each cluster has $O((\log n) / \beta)$ diameter and there are at most $\beta m$ edges connecting distinct clusters. The choice of $\beta$ controls the clustering resolution.
Our implementation is based on the parallel MPX algorithm~\cite{miller2013parallel}.

\myparagraph{$k$-core Decomposition (KCore)~\cite{dhulipala2017julienne, gbbs}}
$k$-core decomposition is a graph clustering technique that recursively prunes vertices from the graph whose degree is less than $k$, until all remaining vertices have at least $k$ neighbors. This process partitions the graph into $k$-cores, where each $k$-core is a maximal subgraph in which every vertex has degree at least $k$ within that subgraph. Non-empty $k$-cores for large value of $k$  tend to represent a densely connected core of the graph, while $k$-cores for small values of $k$ represent more peripheral regions. For a given value of $k$, $k$-core decomposition identifies clusters by treating each maximal $k$-core subgraph as a cluster and treating all other vertices  as singleton clusters. The choice of $k$ can be used to control the clustering resolution.

The algorithm outputs \emph{connected components} of vertices with core number at most $k$.
Vertices with core number less than $k$ are in their own singleton clusters.

\providecommand{\clone}{\overline{N}} %
\providecommand{\set}[1]{\left\{#1\right\}}
\providecommand{\on}[1]{\operatorname{#1}}
\providecommand{\smallabs}[1]{\lvert#1\rvert}

\myparagraph{Structural Graph Clustering (SCAN)~\cite{gbbs,Tseng2021}}
This implementation of SCAN is a parallel version~\cite{Tseng2021} of the index-based SCAN algorithm introduced in \citet{wen2017efficient}. Following~\cite{xu2007scan}, our implementation of SCAN defines a structural similarity $\sigma$ between adjacent vertices $u$ and $v$ as $\sigma(u, v) = 
  \frac{\smallabs{\clone(u) \cap
  \clone(v)}}{\sqrt{\smallabs{\clone(u)}}\sqrt{\smallabs{\clone(v)}}}$, where
$\clone(u)$ is a set including $u$ and the neighbors of $u$.

Two vertices are structurally similar if their structural similarity is at least $\epsilon$, an input parameter. 
A vertex is called a \emph{core} vertex, if it is structurally similar to at least $\mu$ other vertices, where $\mu$ is another parameter.
A cluster is constructed by a core expanding to all vertices that are structurally similar to that core. 

Specifically, the output of the algorithm is equivalent to the following algorithm.
First, construct an auxiliary graph $H$ on the core points, in which we add an edge between any two core vertices, which are structurally similar.
Then, compute the connected components of $H$.
Finally, assign each non-core point which is structurally similar to a core point to the cluster of that core point (in case there is more than one structurally similar core point, one is chosen arbitrarily).
The remaining points are left as singleton clusters.

\myparagraph{Triangle Connected Component Clustering (TECTONIC)}
\parc{} contains a parallel version of the original TECTONIC clustering algorithm~\cite{tsourakakis2017scalable}. The algorithm first weighs the graph edges based on the number of triangles each edge belongs to.
Specifically, if $t(u,v)$ is the number of triangles that edge $(u,v)$ participates in, the weight of an edge $(u,v)$ is set to $\frac{t(u,v)}{\deg(u) + \deg(v)}$.
Then, it removes all edges with weight below a threshold parameter $\theta$, where $\theta$ controls the clustering resolution, and computes connected components.

\myparagraph{Label Propagation (LP)}
\parc{} contains a parallel version of the original label propagation algorithm~\cite{raghavan2007near}. 
The label propagation algorithm works by propagating vertex labels through the graph. Initially, each vertex is assigned a unique label. Then in each iteration, vertices adopt the label that the majority of their neighbors currently have, with ties broken lexicographically. 
This causes label updates to propagate across the graph, with dense regions quickly reaching consensus on a label. All vertices start in their own singleton cluster. The algorithm runs until no vertex updates its label or a maximum number of iterations has been reached. The groups of vertices converging to the same label represent the clusters. In our parallel implementation, on each round, all vertices asynchronously update their labels in parallel. Specifically, on each round, a vertex $v$'s neighbor's label might be updated from $l$ to $l'$ after $v$ has already used the label $l$ to update its own label. 

\myparagraph{Speaker-Listener Label Propagation (SLPA)}
\parc{} contains a parallel version of the original SLPA~\cite{xie2011slpa} algorithm. SLPA is a variant of the label propagation algorithm, where each vertex maintains a list of labels (its memory) instead of a single label. 
On each round, in parallel every vertex 
passes a random label to each neighbor,
chosen from its memory with probability proportional to the occurrence frequency of this
label in the memory (the choice can be different for each neighbor). Whenever a vertex receives a new label from a neighbor, it immediately updates its own memory, so the implementation is asynchronous. 
SLPA may produce overlapping and nested clusters.

\subsection{Scope}\label{sec:scope}
Due to the long history of research on clustering algorithms, hundreds of clustering algorithms have been developed over the past century, and exhaustively comparing all existing proposals is an impossible task.
Instead, we aimed to choose a {\em representative} collection of clustering algorithms that are (1) scalable and efficient and (2) can produce high-precision results.
We now explain our rationale in more detail.

\myparagraph{(1) Scalability and Efficiency} To keep up with the rapid growth of real-world data, a requirement for modern clustering algorithms is that they should be scalable to very large datasets, and scalable with increasing computational power (e.g., cores and machines).
In this paper, we focus on the shared-memory (single-machine) multicore setting, which has been successfully used to process graphs up to hundreds of billions of edges~\cite{gbbs}.

Our focus on scalable and efficient clustering algorithms rules out algorithms such as \revised{graph neural network approaches~\cite{tsitsulin2023graph}, exact spectral clustering~\cite{ng2001spectral, chen2010parallel}} and other algorithms based on numerical linear algebra, whose scalability is limited due to their poor computational efficiency.

\myparagraph{(2) High Precision}
In many applications of clustering, such as spam and abuse detection, classification, and deduplication, a usual requirement is to produce a {\em high precision} clustering.
In other words, it is more important to ensure that all entities within a cluster are closely related, than to ensure that each group of related entities ends up in a single cluster.
Because of this reason we do not include balanced graph partitioning in our comparison, which primarily optimizes recall (i.e., most vertices in each ground truth cluster are grouped into the same output cluster). 
We note that balanced partitioning algorithms are \textit{not} included in many of the popular graph libraries and databases~\cite{staudt2016networkit, neo4j, tigergraph, nebulagraph, memgraph}.

\subsection{Clustering Quality Metrics} \label{sec:metric}
\parc{} includes parallel implementations of many popular metrics for evaluating clustering quality, including
precision, recall, $F$-score, adjusted rand index (ARI), normalized mutual information (NMI), edge density, triangle density, and more.

\myparagraph{Precision, Recall, and $F$-score}
\parc{} computes the average precision and recall of a clustering compared to the ground truth. These metrics work for both non-overlapping and overlapping clusters.
To compute average precision and recall, for each
ground-truth community $c$, we match $c$ to the cluster $c'$ with the
largest intersection to $c$. We list the formulae for computing the metrics below. This metric matches the methodology used by
\citet{tsourakakis2017scalable} in evaluating TECTONIC and \citet{shi2021scalable} in evaluating correlation clustering. 
\vspace{-5pt}
\begin{align*}
\mathrm{Precision} & = |c \cap c'| / |c'| \hspace{30pt}
\mathrm{Recall}   = |c \cap c'| / |c| \\
F_{\beta} &  = (1 + \beta^2) \cdot \frac{\mathrm{Precision} \times \mathrm{Recall}}{(\beta^2 \times \mathrm{Precision}) + \mathrm{Recall}}
\end{align*}
Here, $\beta$ is a parameter specifying the relative importance of precision and recall.
When $\beta < 1$, the objective rewards optimizing precision over recall.
When $\beta > 1$, optimizing recall becomes more important.

\myparagraph{ARI and NMI} 
Both Adjusted Rand Index (ARI)~\cite{hubert1985comparing, chacon2023minimum} and Normalized Mutual Information (NMI)~\cite{danon2005comparing} measure the similarity between two clusterings of a dataset, and are widely used to evaluate the quality of non-overlapping clusters. See \citet{hubert1985comparing} and \citet{danon2005comparing} for a definition of these measures. Extensions to overlapping clusters have also been proposed (e.g., \cite{mcdaid2011normalized, lancichinetti2009detecting}).

\myparagraph{Edge and Triangle Density} These two metrics are not defined with respect to ground-truth clustering but rather measure the structural properties and density characteristics of the identified clusters themselves. Edge density quantifies the internal cohesiveness of a cluster by measuring the ratio of edges in the cluster to the maximum possible number of edges. Similarly, triangle density captures the proportion of triangle subgraphs within a cluster to the maximum possible number of triangles, given the existing wedges. 

We compute a weighted mean of edge density, where the density of each cluster is weighted by its size when computing the mean.
We use the weighting because taking an unweighted average of cluster densities can lead to counter-intuitive results. To illustrate this, consider a clustering with 99 clusters of size 2 and density $1$, and one cluster of size 1 million and density $0.01$.
Then, the average density of the 100 clusters is over $0.99$, even though over $99.99\%$ of vertices are in a cluster of density $0.01$.

\myparagraph{Modularity and Correlation Objective} \parc{} computes the modularity and correlation clustering objectives as described in \Cref{sec:algorithms}.

\myparagraph{Other Cluster Statistics} \parc{} also reports the distribution (minimum, maximum, and mean) of cluster sizes, the diameter of clusters, and the number of clusters.

\subsection{Benchmark}\label{sec:benchmark}
\parc{} has a convenient and flexible framework for benchmarking clustering algorithms.
Listing~\ref{lst:config} gives an example of the configuration file for benchmarking different clustering algorithms. Users can flexibly specify the graphs to benchmark, the number of threads to use, the number of rounds to run, the timeout, and the set of parameters to try for each clustering algorithm. Specifically, for each clustering algorithm, \parc{} tries a Cartesian product of all parameter values. 
\revised{We chose to use a Cartesian product so make benchmarking comprehensive, but users can easily modify \parc{} with a different method to explore the parameter space.}
Besides clustering implementations in \parc{}, \parc{} also supports running many clustering implementations in NetworKit, Neo4j, and TigerGraph.

For each graph, each clusterer, each parameter set, and each round of the experiment, \parc{} outputs a file for the resulting clustering in \texttt{Output} \texttt{directory}.
It also outputs a CSV file including the running time of all runs specified by the configuration file to \texttt{CSV} \texttt{Output} \texttt{directory}. 

\parc{} also supports specifying the ground truth communities and computing statistics of the clustering results, such as precision and recall. Listing~\ref{lst:config_stats} gives an example of the statistics configuration file for computing the quality metrics. This configuration file is used together with the clustering configuration file (e.g., Listing~\ref{lst:config}). 
\texttt{statistics\_config} configures which metrics to compute and the parameters for the metrics. \parc{}'s metrics library outputs:
\begin{itemize}[leftmargin=*]
\item a JSON file with the clustering metrics for each combination of the graph, clusterer, parameter set, and experiment round,
\item a CSV file including the metrics of all runs specified by the clustering configuration file
\end{itemize}

\subsection{Extending \parc{}}

\revised{Our \parc{} design, as described in \Cref{sec:benchmark}, can be extended to include new datasets, algorithms, parameter searches, and more.

To add a new dataset, users can simply provide the graph data in edge list or compressed sparse row format, and add it to the configuration file.For integrating new clustering algorithms, users need only implement a function that takes an input graph and algorithm parameters, and returns a clustering result. This function can then be registered with \parc{}'s benchmarking framework. Users can also extend the parameter search methods by implementing custom search strategies, enabling more efficient exploration of algorithm parameter spaces. New evaluation metrics can also be added in a similar way.

Additionally, PCBS outputs benchmarking results in convenient pandas dataframe and CSV format, which allows for easy integration of visualization tools, or even entire modules for specific analysis tasks. 

By leveraging these extension points, researchers and practitioners can adapt PCBS to their unique requirements, facilitating direct comparisons between novel methods and state-of-the-art algorithms on both standard and domain-specific graphs.

}

{\small
\begin{lstlisting}[float=tp,basicstyle=\footnotesize\ttfamily, breaklines=true, frame=single, caption={Example Clustering Configuration File}, label={lst:config}] 
Input directory: /input_dir/
Output directory: /output_dir/
CSV output directory: /output_dir_csv/
Clusterers: ParHacClusterer; TigerGraphLouvain
Graphs: lj.gbbs.txt; amazon.gbbs.txt
GBBS format: true
Weighted: false
Number of threads: 60
Number of rounds: 1
Timeout: 7h

ParHacClusterer:
  config:
    weight_threshold: 1.0; 0.3
    epsilon: 0.01; 0.1; 1

TigerGraphLouvain:
  config:
    maxIterations: 10; 20
\end{lstlisting}
}
{\small
\begin{lstlisting}[float=tp,basicstyle=\footnotesize\ttfamily, breaklines=true, frame=single, caption={Example Statistics Configuration File}, label={lst:config_stats}]
Input communities: lj.cmty; amazon.cmty

statistics_config:
  compute_edge_density: true
  compute_precision_recall: true
  f_score_param: 0.5
\end{lstlisting}
}

%% file: exp.tex
\section{Empirical Evaluation}\label{sec:exp}
This section presents the results obtained by running the \parc{} benchmark.
We study the performance of all algorithms described in \Cref{sec:algorithms} on a variety of different graphs.

We show that using our \parc{} benchmark library, we are able to obtain new insights on comparing scalable parallel graph clustering algorithms. In our experiments, we mainly want to answer the following questions:

\begin{enumerate}
    \item For the same algorithm, how fast is the implementation in \parc{} compared to the implementations in other state-of-the-art graph libraries and databases? (\Cref{sec:runtime})
    \item How do the quality and running time of the \parc{} implementations compare on different tasks and which implementation is the most suitable for the task? (Sections~\ref{sec:unweighted}--\ref{sec:high_resolution})
    \item How do different implementations that optimize the modularity/correlation objective compare with respect to the ground truth? (\Cref{sec:modularity})
    \iffull
    \item How does the sparsity of the constructed similarity graph impact clustering quality? 
    (\Cref{sec:compare_k})
    \fi
\end{enumerate}

\revised{We evaluate the clustering algorithms on four different tasks: community detection, vector embedding clustering, dense subgraph partitoining, and high resolution clustering.}
For each task, the experiments use different input data and/or objectives. While we try to group the experimental results into tasks to help organize the results and analyze the behavior of the algorithms, we note that the boundary between the tasks is not strict.

\myparagraph{Results Summary} We find that PCBS is consistently much faster than Neo4j, TigerGraph, and SNAP, and in most cases also faster than NetworKit. 
However, it is worth noting that graph databases (e.g. Neo4j and TigerGraph) offer more comprehensive database functionality in addition to clustering.

We also observed that correlation clustering produces top quality clusters in community detection, vector embedding clustering, and dense subgraph partitioning. 
While the quality of correlation clustering has been studied on unweighted graphs (e.g., \cite{shi2021scalable, veldt2018correlation}), to the best of our knowledge, we are the first work to comprehensively evaluate it on weighted graphs.

ParHAC~\cite{dhulipala2022hierarchical} also achieves relatively high quality. While a single run of ParHAC may be slower than other methods, it offers the advantage of being a hierarchical approach. This means that in a single run, it can generate clusterings at multiple levels of granularity. In contrast, methods like correlation and modularity clustering can only produce a single clustering in a one run.

\myparagraph{Parallel Computation Environment}
We use \textit{c2-standard-60} instances on the Google Cloud Platform. These are 30-core machines with two-way hyper-threading with Intel 3.1 GHz Cascade Lake processors (with a maximum turbo clock speed of 3.8 GHz).
We use all 60 hyper-threads for our experiments, except for the experiments specifically investigating how performance scales with the number of threads, and for Neo4j, where we only use 4 threads because that is the maximum number of threads supported in the community version.

\myparagraph{Datasets}
We use a variety of large real-world and synthetic datasets, summarized in \Cref{table:graph_datasets,table:weighted_datasets,table:rmat_datasets}. 
We also present results on five small datasets from UCI machine learning repository~\cite{asuncion2007uci} 
\iffull
in \Cref{sec:uci}. 
\fi
All graphs are undirected.

\Cref{table:graph_datasets} presents unweighted real-world graphs from the Stanford Network Analysis Project (SNAP)~\cite{leskovec2016snap}. For these graphs, we used the top 5000 communities as the ground truth, and the ground truth clusters may be overlapping.

\Cref{table:weighted_datasets} describes the real-world vector embedding datasets~\cite{yu2023pecann} from which we derive our weighted $k$-nearest neighbor graphs.
The embeddings all have 1024 dimensions, except for MNIST, whose dimensionality is 768. The edge weights between vertices $u,v$ are $\frac{1}{1+d(u,v)}$,
where $d(u,v)$ is the Euclidean distance between points $u,v$. The ground truth clusters are non-overlapping. 
\iffull
In this section, we present the results for $k=50$. In \Cref{sec:compare_k}, we also present the results for $k=10$ and $k=100$. 
\fi

\Cref{table:rmat_datasets} presents synthetic unweighted RMAT graphs~\cite{ChakrabartiZF04} generated using GBBS~\cite{gbbs} with parameters $a=0.5$, $b=c=0.1$, and $d=0.3$. One of them is relatively dense (\datasetname{RMAT-1}), while the other is relatively sparse (\datasetname{RMAT-2}).

\revised{We also generated a dataset by extracting 1,274,126 frequently occurring short texts from the NGrams dataset~\cite{Bhatia16}, selecting those that appear at least 120 times. These texts were embedded using the textembedding-gecko@003 model~\cite{gecko}, and an exact 50-nearest-neighbor undirected graph was computed from the embeddings. We created $96\,000$ labels by sampling point pairs across predefined similarity buckets (from 0.76 to 0.99), ensuring diversity in both the data points and their embedding similarities.\footnote{This dataset is available for download at \url{https://storage.googleapis.com/ngrams-similarity-ds/index.html}} Finally, we labeled these pairs based on their embedding similarity, designating pairs with a similarity above 0.92 as belonging to the same cluster and the rest as belonging to different clusters. }

\iffull
More details on these datasets are also presented in \Cref{appendix-dataset}.
\fi

\input{datasets}

\myparagraph{Evaluation}
We evaluate the clustering quality by metrics that compare with the ground truth as well as the edge density of the clusters, which is not associated with any ground truth. 
We use precision, recall, and $F_{\beta}$ score with $\beta = 0.5$ to measure the difference between our clustering and the ground truth.
We note that we use $\beta=0.5$ instead of the default setting of $\beta=1$, which puts more weight on precision than recall.
This is because our focus is on finding clusters of high precision.

We also compute the area under the precision vs.\ recall curve (AUC).
Again, since we focus on the high-precision regime, we only consider results whose precision is in $[0.5, 1]$.
A larger AUC score means the method obtains both high precision and recall.
We report the AUC times 2, so a perfect algorithm gets an AUC of 1. 
We use AUC in addition to $F_{\beta}$ score because it considers the overall performance of an algorithm for many different parameter values, while the $F_{\beta}$ score measures the performance of an algorithm at a single set of parameters.

\myparagraph{Pareto frontier} We present the Pareto frontiers of (a) precision vs.\ recall and (b) clustering quality ($F_{0.5}$ score) vs.\ runtime. 
The Pareto frontier comprises non-dominated points.
To create the Pareto frontier, we run the clustering algorithms with different parameters that allow the resulting clustering to cover the whole precision range as much as possible. 
The result of each run is then represented as a point $(x, y)$, e.g., in the context of clustering quality vs.\ runtime, $x$ is the running time and $y$ is the $F_\beta$ score. 
After that, we keep each point $(x, y)$, for which  there does not exist another point $(x', y')$, such that $x' \geq x$ and $y' \geq y$.
In the precision vs.\ recall Pareto frontier plots, the \emph{top right} is better (high precision and high recall). In the clustering quality vs.\ runtime Pareto frontier plots, the \emph{top left} is better (high quality and low running time). %

\begin{figure}[t]
    \centering
    \includegraphics[width=\columnwidth]{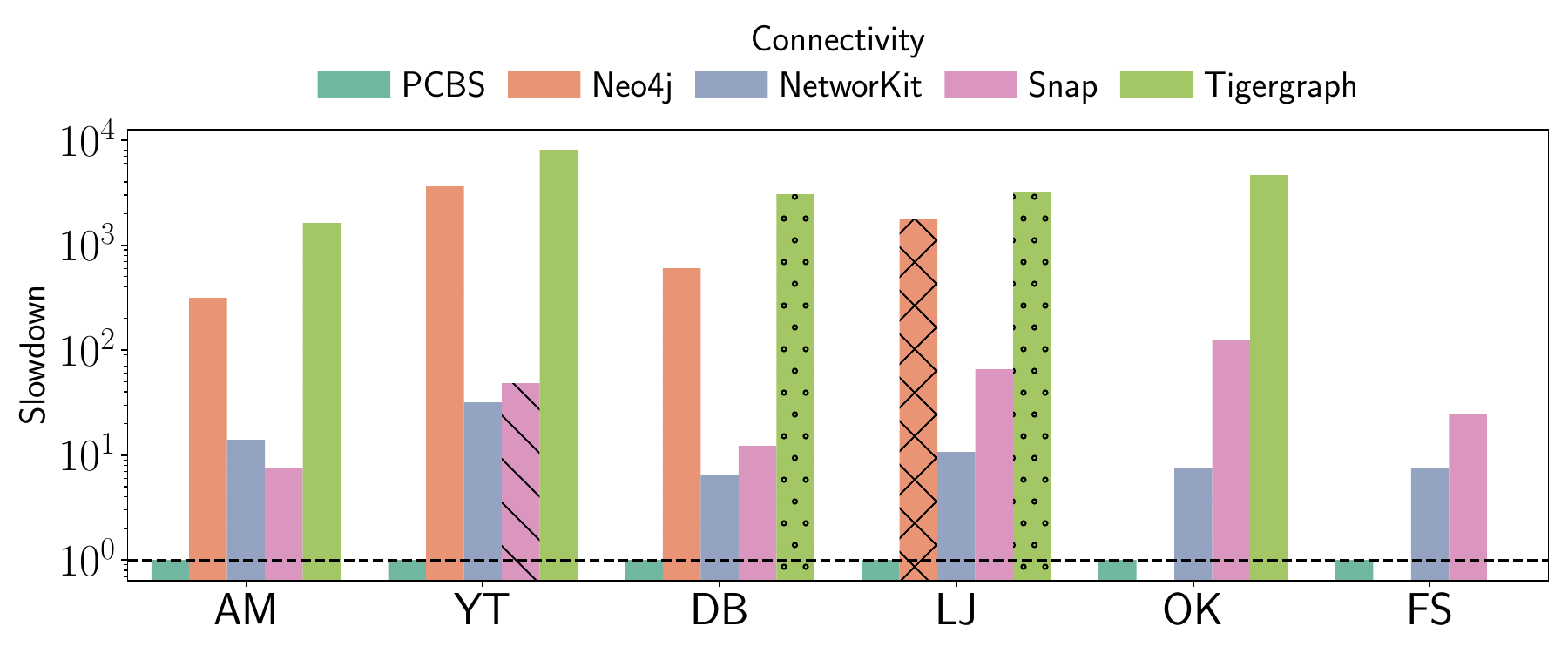}
    \vspace{-1.5em}
    \vspace{-1em}
    \includegraphics[width=\columnwidth]{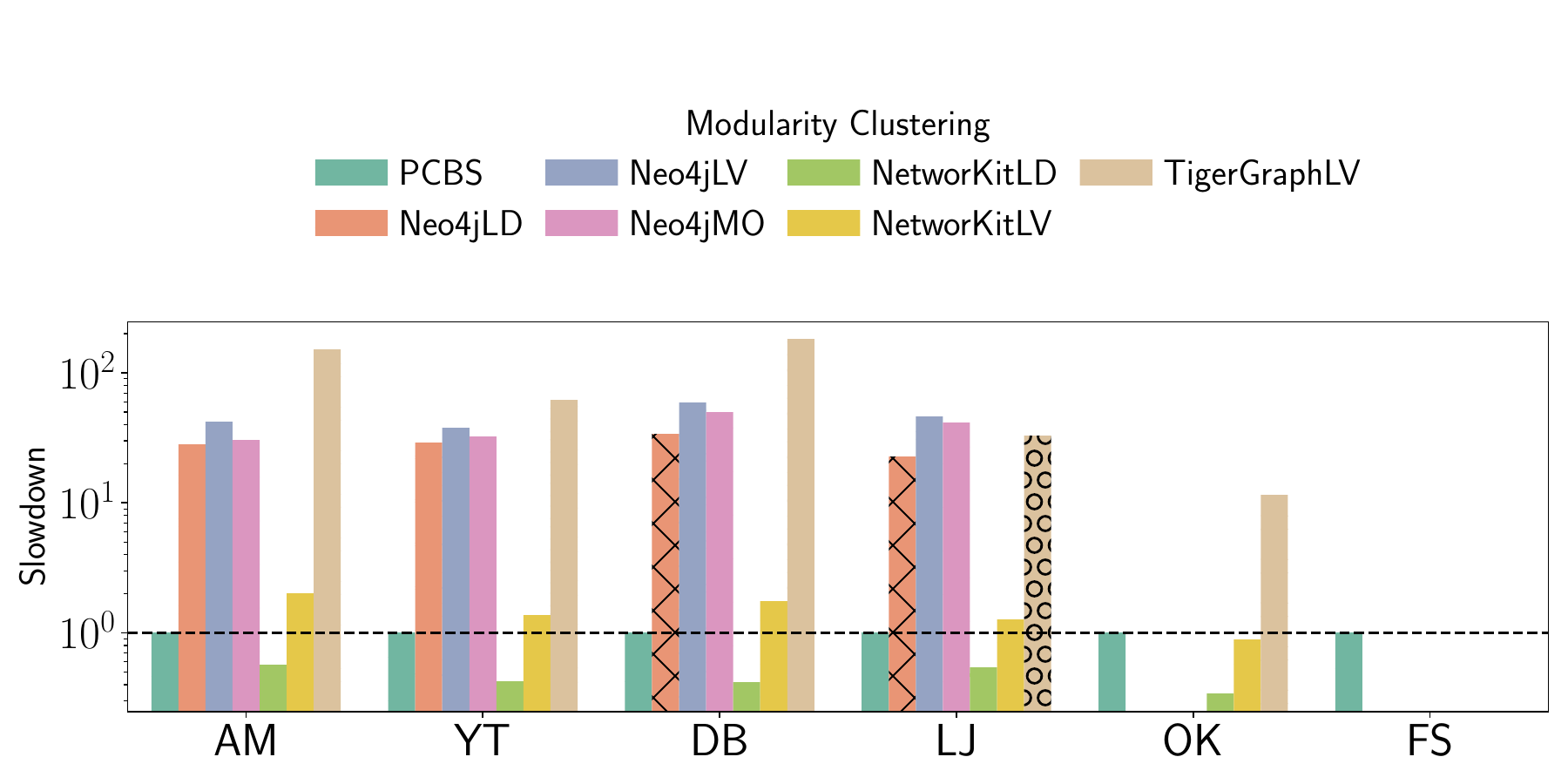}
    \vspace{-2em}
    \caption{Slowdown of methods on SNAP graphs with respect to \parc{}. Neo4j cannot load \datasetname{orkut} and \datasetname{friendster}. TigerGraph cannot load \datasetname{friendster}. NetworKit failed to run on \datasetname{friendster}.
    "LD" methods are Leiden-based. "LV" methods are Louvain-based. "MO" is a Girvan-Newman implementation. 
    The horizontal dashed line is at slowdown=1.
    }
    \label{fig:slowdown}
\end{figure}

\input{table_platforms}

\myparagraph{Baseline Graph Libraries and Databases}
We compare with NetworKit~\cite{staudt2016networkit}, Neo4j~\cite{neo4j}, TigerGraph~\cite{tigergraph}, the original implementation of TECTONIC~\cite{tsourakakis2017scalable}, and clustering implementations provided in SNAP~\cite{leskovec2016snap}. All algorithms are parallel, except for  the original implementation of TECTONIC and the implementations in SNAP, which are sequential. The list of algorithms implemented in each graph library or database can be found in \Cref{tab:all_platforms}. 

The NetworKit algorithms are implemented in C++ with Python bindings. Neo4j's core components are implemented in Java. TigerGraph algorithms are implemented using GSQL on top of the core components implemented in C++.

The original TECTONIC implementation counts triangles in C, and then writes the results to disk. Finally, C++ is used to read the triangles from the disk and compute the clusters. It does not appear to be optimized for running time.

Several graph libraries and databases have multiple different modularity clustering implementations, and we describe them here. Our parallel modularity and correlation clustering implementation is Louvain-based and optimizes the LambdaCC objective, as described in \Cref{sec:algorithms}.

\begin{itemize}[leftmargin=*]
    \item NetworKit~\cite{staudt2016networkit} has two parallel implementations for modularity clustering: Louvain method (NetworKitPLM)~\cite{Staudt2016plm} and Leiden Method~\cite{leiden} (NetworKitParalleLeiden).
    \item SNAP~\cite{leskovec2016snap} has two sequential implementations: CommunityGirvanNewman~\cite{GiNe02} and CommunityCNM~\cite{clauset2004finding}. 
    \item Neo4j~\cite{neo4j} has three parallel implementations: Louvain method (Neo4jLouvain), Leiden method (Neo4jLeiden), and a Girvan-Newman method (Neo4jModularityOptimziation).  
    \item TigerGraph~\cite{tigergraph} has a parallel Louvain clustering implementation (TigerGraphLouvain). 
\end{itemize}
We compare these modularity implementations in \Cref{sec:runtime} and \Cref{sec:modularity}. Other algorithms all have a single implementation in each library/database, and we compare them in  \Cref{sec:runtime}.

\subsection{Running Time Comparison with Baselines}\label{sec:runtime}
In \Cref{fig:slowdown}, we show the running time comparison of \parc{} and the baselines on selected clustering methods. The $y$-axis is the slowdown compared to the method in \parc{}. Other methods show a similar trend, and we present the running time comparison of all methods and their scalability 
\iffull
in \Cref{sec:runtime-appendix} (\Cref{fig:scalability}).
\fi

On all tested implementations, \parc{} is more than an order of magnitude faster than Neo4j, TigerGraph, and SNAP. 
\iffull
Although SNAP and Neo4j do not use all 60 threads, we show in \Cref{fig:scalability} that when using the same number of threads, \parc{} is still significantly faster.
\fi

We attribute the performance of \parc{} to three levels of optimization: (1) on the algorithm level, it uses theoretically-efficient algorithms; (2) on the implementation level, it is carefully engineered by, e.g., utilizing cache efficiently and minimizing the use of locks to reduce contention; and (3) on the system level, it uses an efficient C++ framework for parallel graph algorithms, i.e., GBBS~\cite{gbbs} and ParlayLib~\cite{parlaylib}.
\revised{While the Parlay and GBBS libraries contribute to the speedup by providing essential primitives for working with sequences and graphs, most of the overall speedup comes from algorithm-specific optimizations. 
We note that primitives in Parlay have similar speed as counterparts in other parallel runtimes like TBB and ParallelSTL~\cite{parlaylib}.
The remaining optimizations are generally algorithm-specific, and so for illustration we provide a more comprehensive discussion of one algorithm later in the section. For the remaining ones, we refer readers to the respective clustering algorithm papers for further insights.}

Compared to NetworKit, \parc{} is significantly faster on the connectivity clustering and $k$-core clustering tasks. On modularity clustering, \parc{}'s running time is comparable to NetworKitPLM (NetworKitLV), but slower than NetworKitParallelLeiden (NetworKitLD). However, in our experiments, NetworKit failed to run on \datasetname{friendster}, and our implementation is the only implementation that successfully ran on \datasetname{friendster}.
Contrary to connectivity and k-core, these modularity clustering implementations can produce different clusterings, we further study the running time and quality of these different modularity clustering implementations under different parameters in \Cref{sec:modularity}. Though NetworKitParallelLeiden is faster, we will see that it obtains lower quality than \parc{}'s modularity clustering on some data sets.

On the largest two graphs, \datasetname{orkut} and \datasetname{friendster}, some bars are missing because Neo4j cannot load \datasetname{orkut} and \datasetname{friendster}, and TigerGraph cannot load \datasetname{friendster} due to memory constraints. The SNAP modularity clustering implementation timed out on all graphs with a time limit of 7 hours.
We suspect that slower performance of Neo4j and Tigergraph compared to \parc{} and NetworKit comes from the fact that they support a more general graph database functionality, which incurs additional overheads.

We also compared our parallel TECTONIC implementation with the original TECTONIC~\cite{tsourakakis2017scalable} implementation, which is sequential. 
On the \datasetname{youtube} data set with threshold 0.06, our implementation is 37x faster than the original TECTONIC when run on a single thread, and 387x faster when run on 60 threads.

\begin{figure}[t]
    \centering
    \includegraphics[width=0.8\columnwidth]{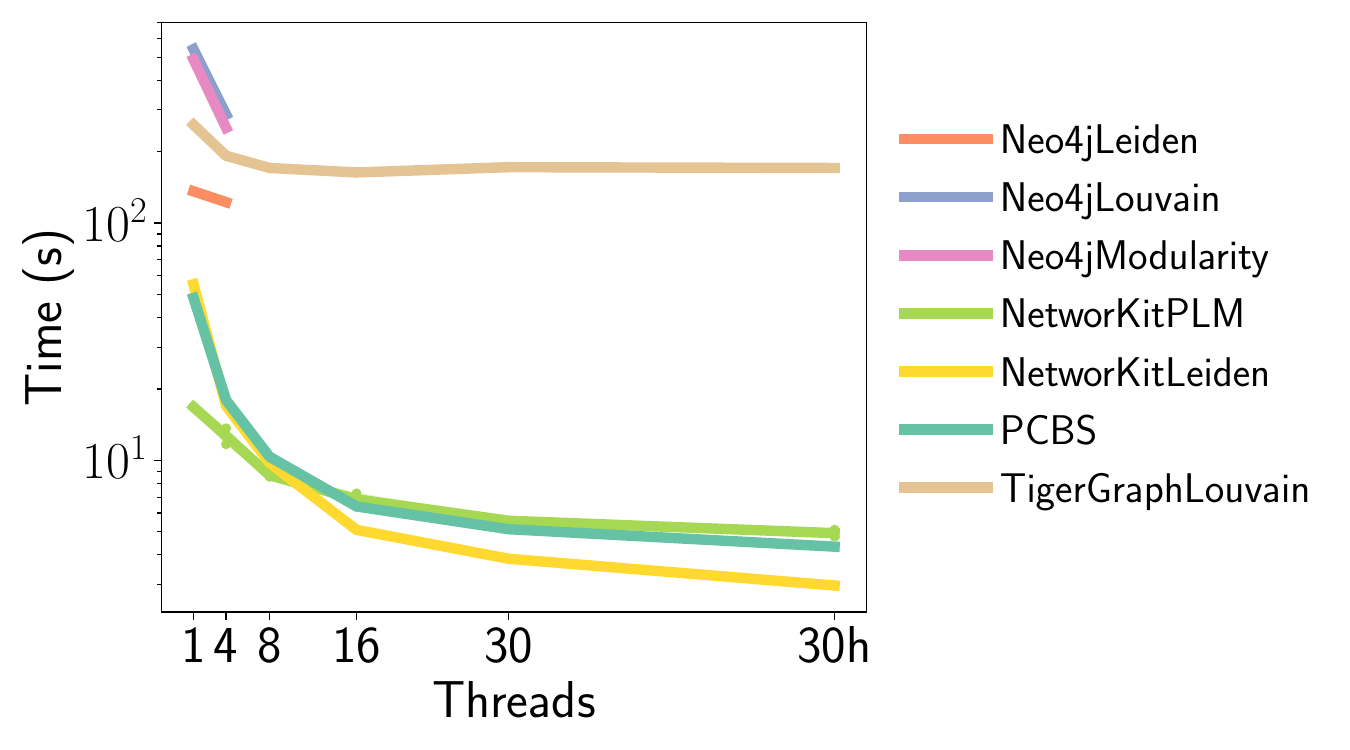}
    \vspace{-1.5em}
    \caption{Scalability of modularity clustering implementations on \datasetname{lj} with a resolution parameter of 1 and a maximum iteration count of 10. `30h' means using all 30-cores with two-way hyperthreading.}
    \label{fig:scalability-md}
\end{figure}

For KCore, connectivity, TECTONIC, Label Propagation, and SLPA, we use our implementation in the rest of the experiments because \parc{}'s implementations are the fastest.
For KCore, connectivity, and TECTONIC,
\parc{}'s implementations compute the same clustering as baseline implementation. 
In the case of Label propagation and SLPA, while the algorithms are essentially the same, the resulting clusterings can be different because of non-determinism in thread scheduling and randomization.

In \Cref{fig:scalability-md}, we show the running time of different modularity clustering implementations using various numbers of threads. We see that \parc{} is faster than Neo4j and TigerGraph on all thread counts. 
On lower thread counts, NetworKitPLM (NetworKitLV) is faster than \parc{} and NetworKitParallelLeiden (NetworKitLD), but for more than 8 threads, NetworKitParallelLeiden (NetworKitLD) is the fastest. Though NetworKitLD is the fastest, its clustering quality is on average worse than \parc{}'s correlation clustering as shown in \Cref{sec:modularity}.

\revised{Our correlation clustering implementation employs several key optimizations. It uses an asynchronous, lock-free approach for local search, addressing symmetry breaking challenges in parallel execution. We implement a heuristic to skip unnecessary re-evaluations of vertex movements, improving efficiency. Additionally, our implementation parallelizes the graph contraction step, which is often executed sequentially in other implementations. These optimizations collectively contribute to the superior performance of our correlation clustering algorithm.}

While graph databases, such as Neo4j~\cite{neo4j} and TigerGraph~\cite{tigergraph}, provide a richer functionality, on different graph clustering implementations, they are slower. For example, on the LiveJournal graph from SNAP~\cite{leskovec2016snap}, \parc{} is on average 32.5x faster than Neo4j and 303x faster than TigerGraph.
Compared with state-of-the-art parallel graph library NetworKit~\cite{staudt2016networkit}, \parc{} is on average 4.54x faster. We compute the average using the geometric mean.
For Neo4j, we use the execution time with 4 threads for both Neo4j and PBCS when computing the average speedup.
We considered the following algorithms:
connectivity, KCore, Neo4jLeiden, Label Propagation, and SLPA.
For TigerGraph and NetworKit, we use the execution time with 60 threads for TigerGraph, NetworKit, and PBCS.
We considered the following algorithms:
connectivity, KCore, TigerGraphLouvain/NetworKitParallelLeiden, and Label Propagation.

\subsection{Community Detection}\label{sec:unweighted}
In this section, we compare the algorithms on a community detection task.
Here the input is an unweighted graph, whose edges describe real-world relations between individuals.
The goal is to recover ground truth communities in the graph with high precision, by clustering the graph vertices.

\begin{figure}[t]
    \centering
    \includegraphics[width=\columnwidth]{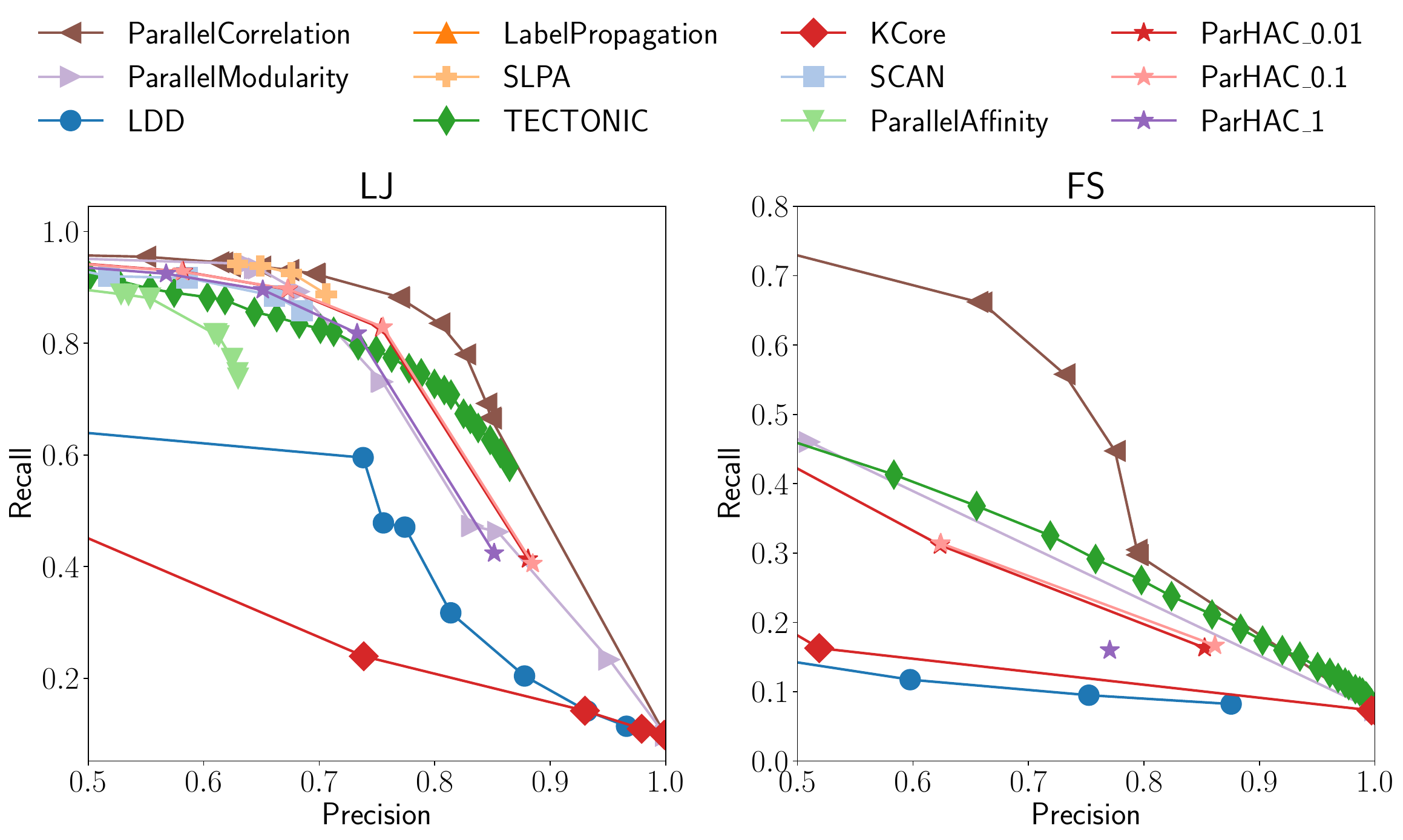}
    \includegraphics[width=\columnwidth]{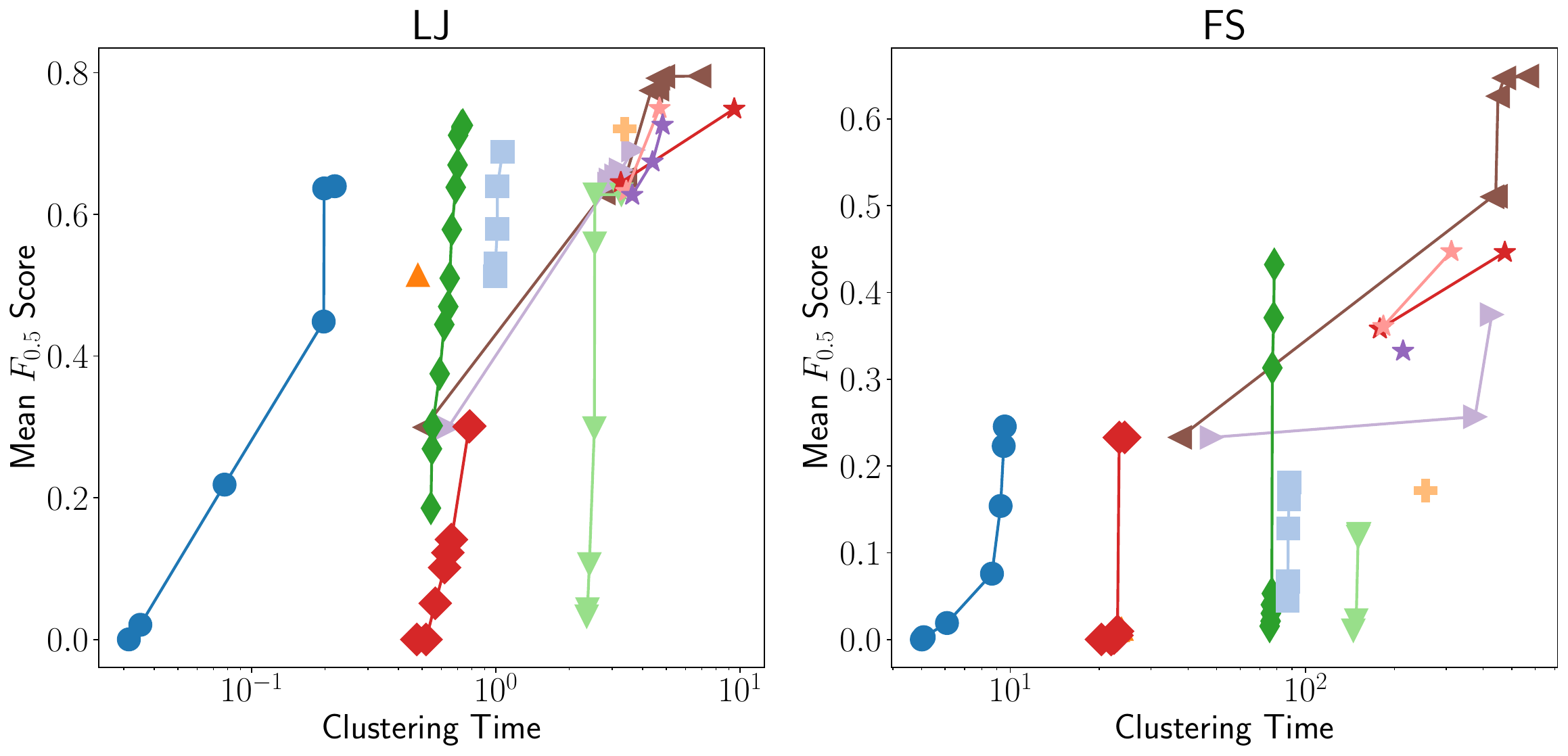}
    \vspace{-2em}
    \caption{\textbf{(Top)} The Pareto frontier of the precision and recall of the unweighted SNAP graphs. \textbf{(Bottom)} The Pareto frontier of the $F_{0.5}$ and runtime graph for the unweighted SNAP graphs. "ParHAC$\_{\epsilon}$" shows the curve for ParHAC implementation with approximation parameter $\epsilon$. 
    }
    \label{fig:pr_snap_subset}
\end{figure}

 \begin{table}[t]
 \small
\centering
\caption{Area under precision-recall curve (AUC) for precision $\geq$ 0.5 on SNAP graphs. }
\vspace{-1em}
    \label{tab:snap}
\begin{tabular}{l|ccccccc}
\toprule
Clusterer & LJ & AM & DB & YT & OK & FS & Mean \\
\midrule
Correlation & \textbf{ 0.75 } & \textbf{ 0.94 } & \textbf{ 0.71 } & 0.41 & \textbf{ 0.32 } & \textbf{ 0.43 } & \textbf{ 0.59 } \\
TECTONIC & 0.67 & 0.94 & 0.67 & 0.42 & 0.21 & 0.29 & 0.53 \\
Modularity & 0.66 & 0.92 & 0.63 & \textbf{ 0.46 } & 0.18 & 0.27 & 0.52 \\
ParHAC-0.1 & 0.67 & 0.93 & 0.66 & 0.31 & 0.15 & 0.22 & 0.49 \\
ParHAC-0.01 & 0.66 & 0.93 & 0.66 & 0.30 & 0.15 & 0.21 & 0.49 \\
ParHAC-1 & 0.63 & 0.92 & 0.61 & 0.27 & 0.10 & 0.12 & 0.44 \\
SLPA & 0.65 & 0.86 & 0.57 & 0.00 & 0.00 & 0.00 & 0.35 \\
SCAN & 0.61 & 0.87 & 0.51 & 0.00 & 0.00 & 0.00 & 0.33 \\
Affinity & 0.49 & 0.81 & 0.47 & 0.00 & 0.00 & 0.00 & 0.30 \\
LDD & 0.42 & 0.60 & 0.38 & 0.25 & 0.00 & 0.09 & 0.29 \\
KCore & 0.20 & 0.55 & 0.23 & 0.39 & 0.03 & 0.12 & 0.25 \\
LP & 0.00 & 0.84 & 0.44 & 0.00 & 0.00 & 0.00 & 0.21 \\
\bottomrule
\end{tabular}

\end{table}

In \Cref{fig:pr_snap_subset}, we show the Pareto frontier plots on four unweighted SNAP graphs.
The results for all six graphs can be found in
\iffull
 \Cref{fig:pr_snap,fig:time_f1_snap} in the Appendix~\ref{sec:snap_full} 
\fi
(the two other graphs show a similar trend). 
We also present the area under the precision-recall curve (AUC) in \Cref{tab:snap}. 
We exclude connectivity because for unweighted graphs, all vertices are in a single connected component.

Overall, correlation clustering usually achieves the best quality (for both $F_{0.5}$ score and AUC score), but it is relatively slow compared to other clustering implementations. TECTONIC usually achieves lower quality than correlation clustering, but it is much faster. LDD is the fastest method but has even lower quality than TECTONIC. 
TECTONIC often has higher quality than other methods such as LDD, KCore, LP, SCAN, and Affinity.

On the largest two graphs, \datasetname{orkut} and \datasetname{friendster}, correlation clustering gets a significantly higher $F_{0.5}$ score than the other methods.
Modularity clustering obtains a lower score than correlation clustering on \datasetname{friendster}, because for large threshold parameters which correspond to the highest $F_{0.5}$ scores, modularity clustering runs out of memory. In comparison, correlation clustering can run for parameter values spanning the entire precision-recall value range.
ParHAC and affinity overall perform worse than correlation clustering, which is likely due to the fact that they heavily rely on edge weights (which in this task are all set to 1).
On \datasetname{friendster}, ParHAC with $\epsilon=1$ also runs out of memory for some threshold values due to the high memory cost of performing a large set of changes to the contracted graph in the implementation when using large values of $\epsilon$.

Compared to other methods, LDD and KCore do not have high quality.
SLPA, LP, and SCAN can achieve good quality on the smallest \datasetname{amazon} dataset, but have poor quality on the larger graphs, and have a hard time achieving high precision. These methods do not have a threshold parameter that controls the clustering resolution and so their Pareto frontier cannot span the entire precision range.

Overall, based on our findings, we make the following recommendations for community detection:
\begin{itemize}[leftmargin=*]
    \item For generating high-quality clusters, correlation clustering is recommended. 
    \item If one wants to sacrifice some quality for faster running time, TECTONIC is a good option.
    \item If speed is the top priority, LDD can be used. However, this comes at a cost of a significant drop in quality.
\end{itemize}

\subsection{Vector Embedding Clustering}\label{sec:weighted}

In this task, the goal is to classify data points represented by vector embeddings.
We solve this task by
building a weighted $k$-nearest neighbor graph on the input vectors, i.e., a graph in which each vertex represents a single vector, and each vector is connected to its $k$ nearest vectors.
We use the embeddings described in \Cref{table:weighted_datasets} as input.
Clustering the resulting $k$-nearest neighbor graph is an unsupervised method for classifying the vectors.
While clustering by itself would not produce a label for each cluster, it is a powerful method when the number of classes is not known upfront.

\begin{figure}
    \centering
    \includegraphics[width=\columnwidth]{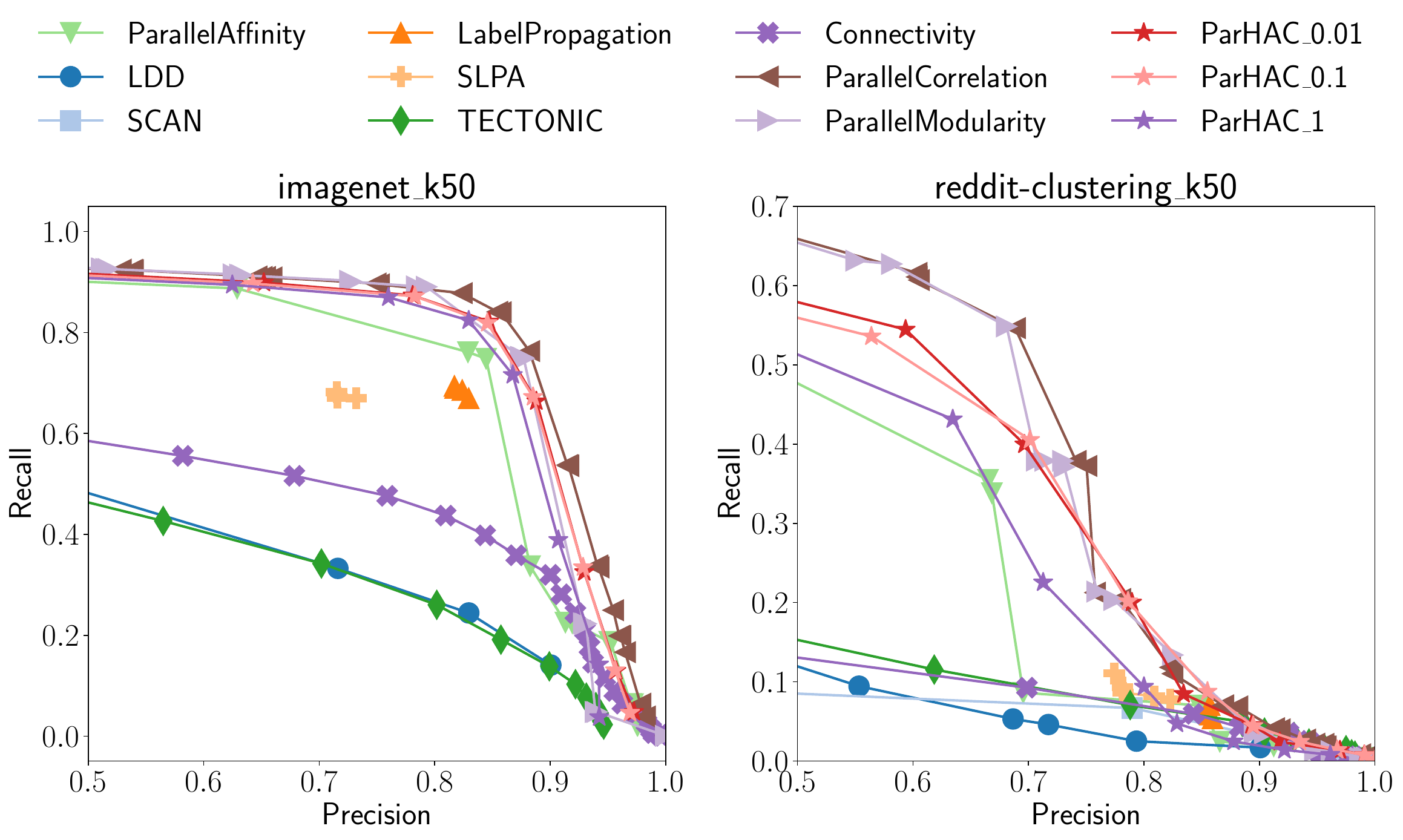}
    \includegraphics[width=\columnwidth]{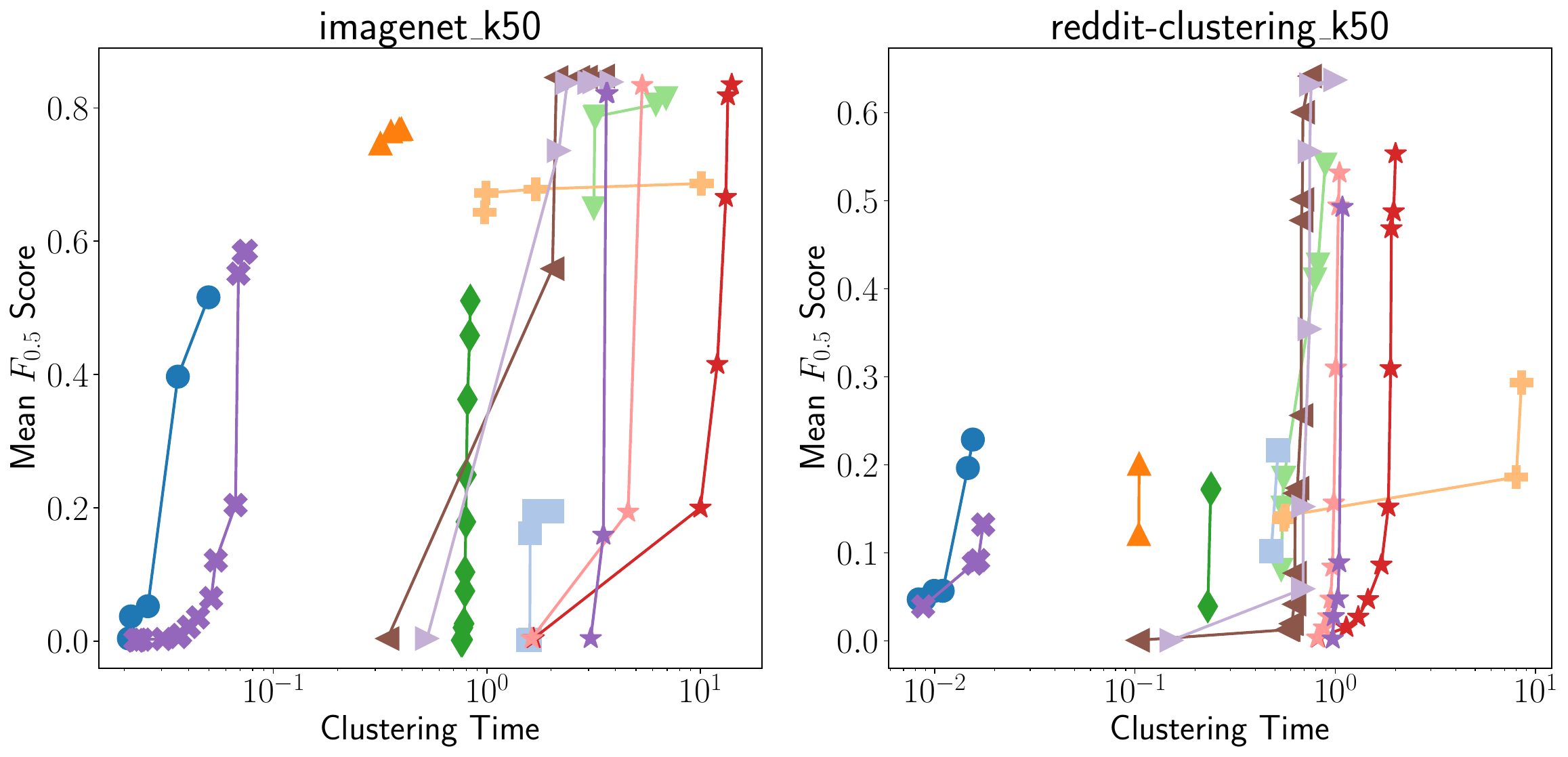}
    \vspace{-2.5em}
    \caption{\textbf{(Top)} The Pareto frontier of precision and recall for the weighted \knn graphs ($k=50$), using \parc{} methods. \textbf{(Bottom)} The Pareto frontier of $F_{0.5}$ score and clustering time on \knn graphs ($k=50$).
    \revised{The plot for all 4 graphs are in 
    \iffull
    \Cref{sec:snap_full} (\Cref{fig:pr_weighted_full}).
    \fi
    }
    }
    \label{fig:pr_weighted}
\end{figure}

We show Pareto frontier plots in \Cref{fig:pr_weighted} and the area under the precision-recall curve (AUC) in \Cref{tab:weighted}. We exclude the $k$-core algorithm because in the $k$-nearest neighbor graphs, all vertices are in the $k$-core.

Here we can see that clustering algorithms that leverage edge weights (affinity, correlation, modularity, and ParHAC) do better than clustering algorithms that do not (LDD, SCAN, LP, SLPA, and TECTONIC). 
One exception is that the simple connectivity algorithm, which leverages edge weights by thresholding on them, is not always better than the algorithms that do not use edge weights.
For example, on \datasetname{ImageNet}, connectivity is noticeably better than LDD and TECTONIC, but it is worse than TECTONIC on \datasetname{MNIST}.

Among the top-performing methods (affinity, correlation, modularity, and ParHAC), correlation clustering and modularity clustering are the fastest and achieve the highest $F_{0.5}$ scores and AUC scores. 
The quality of affinity clustering and ParHAC are comparable and are slightly lower than correlation clustering and modularity clustering on \datasetname{Reddit} and \datasetname{StackExchange}.

The propagation-based methods LP and SLPA can sometimes perform well (e.g., on \datasetname{ImageNet}), but their performance is not stable and they have very low scores on \datasetname{Reddit}. While LP is faster than the top-performing methods, SLPA is not significantly faster and does not show a clear advantage over other methods.

Overall, based on our findings, we make the following observations for the vector embedding clustering task:

\begin{itemize}[leftmargin=*]
    \item For generating high-quality clusters at only a few granularities, correlation clustering and modularity clustering are recommended. 
    \item For generating high-quality clusters at many different granularity levels, ParHAC and affinity clustering are recommended because of their hierarchical nature. Though affinity clustering is slower than ParHAC and correlation clustering on these four data sets, we note that it is faster on larger data sets, as shown in \Cref{sec:unweighted}.
    \item For fast clustering without the need to tune parameters, LP is recommended, although the clustering quality can be unstable.
    \item LDD and connectivity are orders of magnitude faster than correlation clustering. While their quality is significantly lower, the fact that non-trivial clustering can be achieved extremely quickly raises a question whether a fast high-quality method exists.
    \item TECTONIC provides a middle ground in the runtime-quality trade-off. It is faster than correlation clustering, and has higher quality than LDD and connectivity. However, its quality is lower than that of correlation clustering in most cases.
\end{itemize}

\begin{table}[t]
\small
\centering
    \caption{Area under curve for precision $\geq$ 0.5 on weighted $k$-nearest neighbor graphs with $k=50$.}
    \vspace{-1em}
\begin{tabular}{l|lcccc}
\toprule
 Clusterer & MNIST & ImageNet & Reddit & StackExchange & Mean \\
\midrule
Correlation & \textbf{0.88 } & \textbf{ 0.77 } & \textbf{ 0.33 } & \textbf{ 0.20 } & \textbf{ 0.54 } \\
Modularity & 0.87 & 0.73 & 0.32 & 0.19 & 0.53 \\
ParHAC-0.01 & 0.87 & 0.73 & 0.28 & 0.18 & 0.51 \\
ParHAC-0.1 & 0.83 & 0.73 & 0.28 & 0.17 & 0.50 \\
ParHAC-1 & 0.73 & 0.70 & 0.21 & 0.11 & 0.44 \\
Affinity & 0.79 & 0.66 & 0.16 & 0.08 & 0.42 \\
LP & 0.64 & 0.57 & 0.06 & 0.16 & 0.36 \\
SLPA & 0.25 & 0.50 & 0.08 & 0.17 & 0.25 \\
TECTONIC & 0.34 & 0.27 & 0.08 & 0.06 & 0.18 \\
Connectivity & 0.16 & 0.42 & 0.07 & 0.04 & 0.17 \\
LDD & 0.11 & 0.25 & 0.04 & 0.02 & 0.11 \\
SCAN & 0.00 & 0.00 & 0.05 & 0.00 & 0.01 \\
\bottomrule
\end{tabular}

    \label{tab:weighted}
\end{table}

\subsection{Dense Subgraph Partitioning}\label{sec:density}
\begin{figure}
    \centering
    \includegraphics[width=\columnwidth]{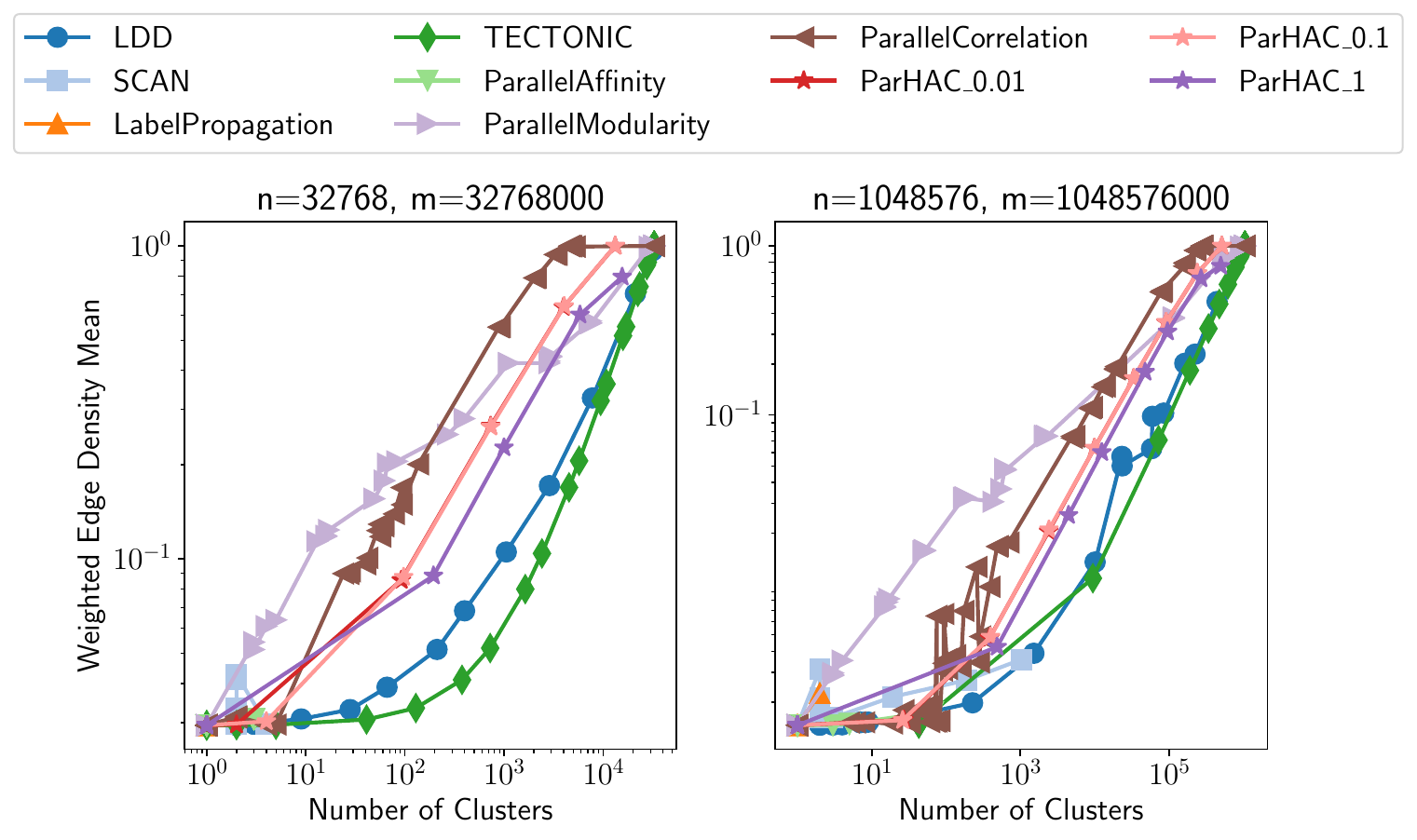}
    \vspace{-2.5em}
    \caption{Weighted average edge density of the clusters in RMAT graphs, with different numbers of vertices ($n$) and edges $(m$).}
    \label{fig:rmat}
\end{figure}

In this section, we evaluate the algorithms on a dense subgraph partitioning task.
Here, the goal is to group vertices into dense clusters.
\revised{Note that this is different from the dense subgraph discovery setting, where we only want to find a single dense subgraph.}
The main difference from the community detection task is that our goal is to directly optimize cluster density, rather than optimizing quality with respect to ground truth labels.
In \Cref{fig:rmat}, we show the {\em weighted edge density mean} of the clusters on artificially generated unweighted RMAT graphs. We refer the readers to \Cref{sec:metric} for the definition of weighted edge density.

Modularity and correlation clustering obtain the densest clusters. When the number of clusters is relatively small, modularity clustering produce clusters denser than correlation clustering on average, but when the number of clusters is very large, correlation clustering produces denser clusters. 
ParHAC also produces dense clusters, but overall less dense than modularity and correlation clustering. LDD and TECTONIC produce less dense clusters. SCAN can produce dense clusters when the number of clusters is very small, but cannot produce dense clusters with larger number of clusters. Label propagation and affinity clustering can only produce small number of clusters, and the density on these two datasets are similar to SCAN.

Overall, we recommend using modularity clustering when the number of clusters is relatively small, and correlation clustering otherwise.

\subsection{High Resolution Clustering}\label{sec:high_resolution}

\begin{figure}[t]
    \centering
    \includegraphics[width=\columnwidth]{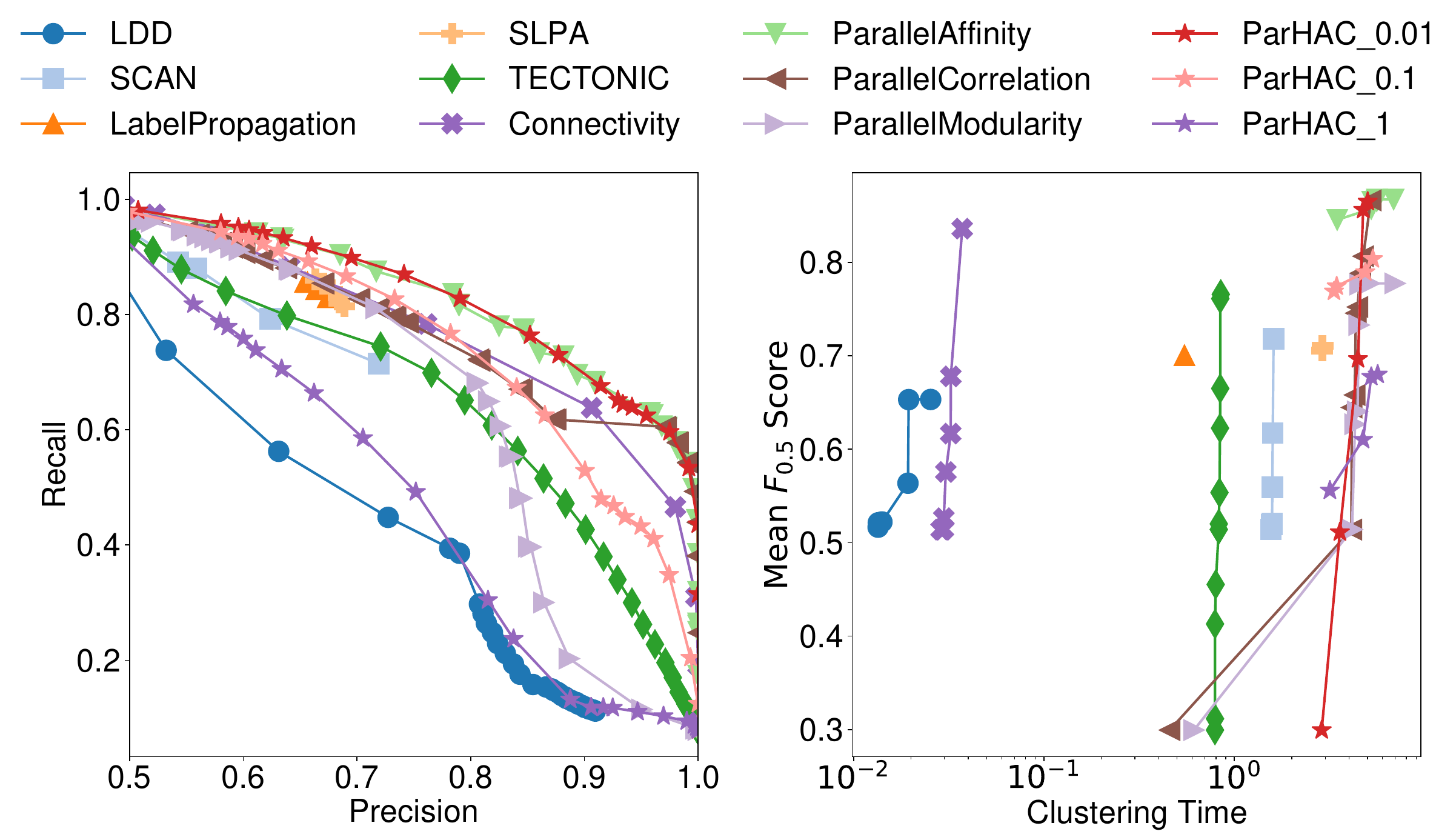}
    \vspace{-2.5em}
    \caption{\revised{\textbf{(Left)} The Pareto frontier of the precision and recall of the NGram graph. \textbf{(Right)} The Pareto frontier of the $F_{0.5}$ score and runtime for the NGram graph. "ParHAC$\_{\epsilon}$" shows the curve for the ParHAC implementation with approximation parameter $\epsilon$. }
    }
    \label{fig:pr_ngram_0.92}
\end{figure}

\revised{This task mimics near-duplicate detection, where we want to find data points that are highly similar to each other. This is a natural task to solve on very large corpora to identify semantically similar entities.
In this task, the ground truth clusters are very small and there are many more ground truth clusters than in the previous tasks. Thus, we call this a \emph{high-resolution clustering} task.
We provide the new NGrams dataset since there are very few large near-duplicate detection datasets with ground truth labels.
\Cref{fig:pr_ngram_0.92} shows the Pareto frontier plots for precision vs.\ recall and $F_{0.5}$ score vs.\ runtime on the NGram graph. We also present the AUC values in 
\Cref{tab:ngram}. 

Overall, ParHAC with small $\epsilon$ values (0.01) and affinity clustering achieve the highest quality, obtaining the best AUC and $F_{0.5}$ scores. We observe that ParHAC's performance decreases rapidly as $\epsilon$ increases, with the $\epsilon=1$ variant performing significantly worse than the variants with smaller $\epsilon$ values. This highlights the importance of using small approximation factors for ParHAC on this type of data, which is expected because we need high accuracy to detect near duplicates.
Correlation clustering and connectivity clustering, while not the top performers, still show good results and are among the better-performing methods.

Based on our findings, we make the following observations:

\begin{itemize}[leftmargin=*]
\item ParHAC with small $\epsilon$ values or affinity clustering are best suited for this task.
\item Connectivity gives close to the best clusters with much faster running time.
\end{itemize}

}

\begin{table}[]
\caption{\revised{Area under curve for precision $\geq 0.5$ on the NGram graph with $k=50$. 
}
}
\vspace{-1em}
    \centering
    \small
\begin{tabular}{lr}
\toprule
ParHAC-0.01  &   \textbf{0.83} \\
Affinity     &              0.82 \\
Correlation  &              0.77 \\
Connectivity &              0.77 \\
ParHAC-0.1   &              0.74 \\
TECTONIC     &              0.64 \\
Modularity   &              0.63 \\
SLPA         &              0.58 \\
LP           &              0.57 \\
SCAN         &              0.56 \\
ParHAC-1     &              0.46 \\
LDD          &              0.38 \\
\bottomrule
\end{tabular}
    \label{tab:ngram}
\end{table}

\subsection{Comparing Different Modularity Clustering Implementations}\label{sec:modularity}

Correlation and modularity clustering methods have shown a strong performance overall compared with other clustering algorithms in our benchmark. 
While parallel modularity clustering algorithms usually follow the same overall framework of local search and graph coarsening~\cite{louvain}, the available implementations differ in a number of ways (e.g., in terms of synchronization, symmetry breaking, heuristic optimizations, or the use of refinement~\cite{shi2021scalable}) that result in differences in the running time and clustering quality. In this section, we perform a comparison between \parc{}'s correlation and modularity clustering and the different baseline clustering algorithms that optimize the modularity objective.

\begin{figure}[t]
    \centering
    \includegraphics[width=\columnwidth, trim={0 0 0 2.6cm},clip]{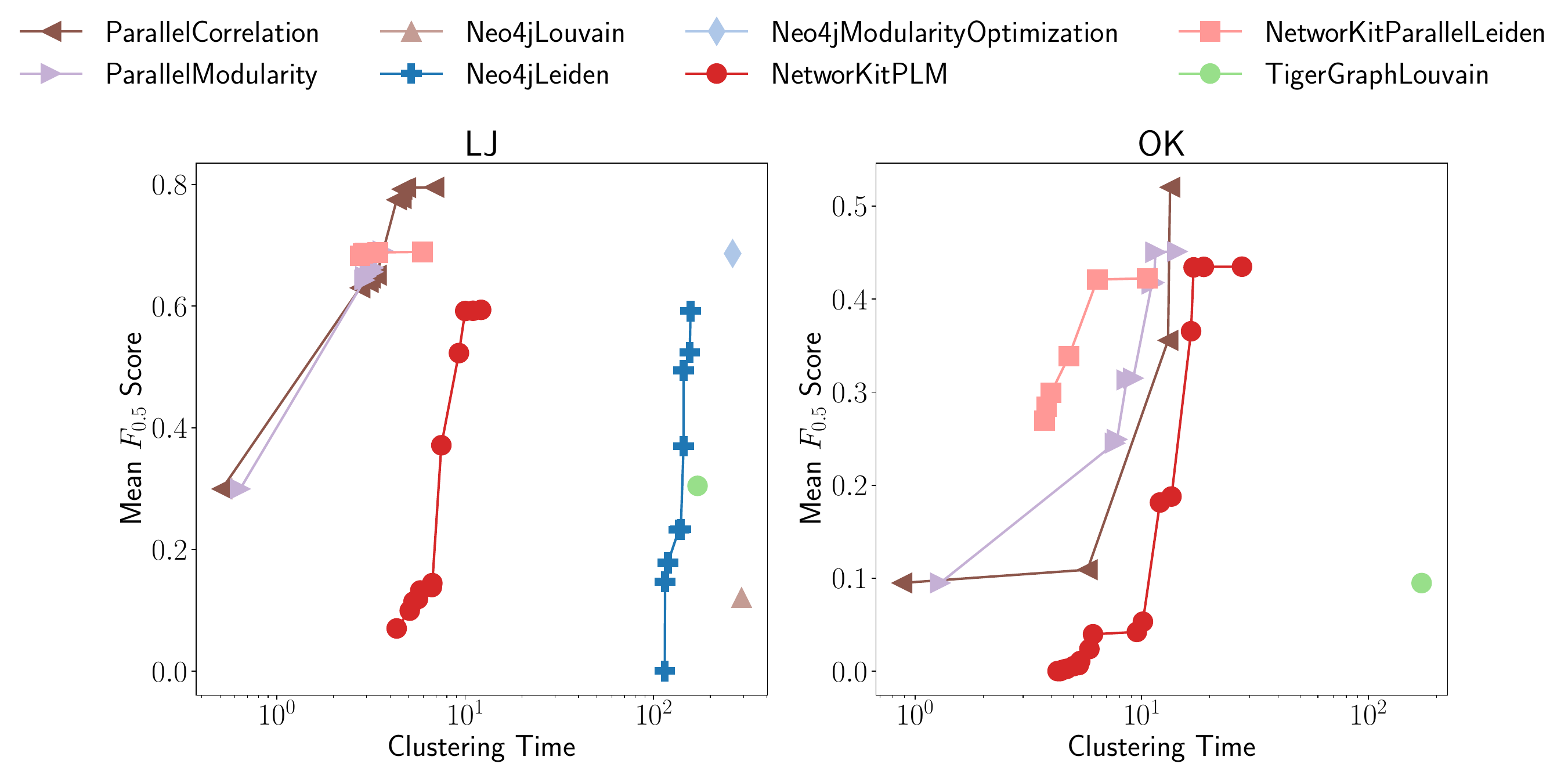}
    \includegraphics[width=\columnwidth]{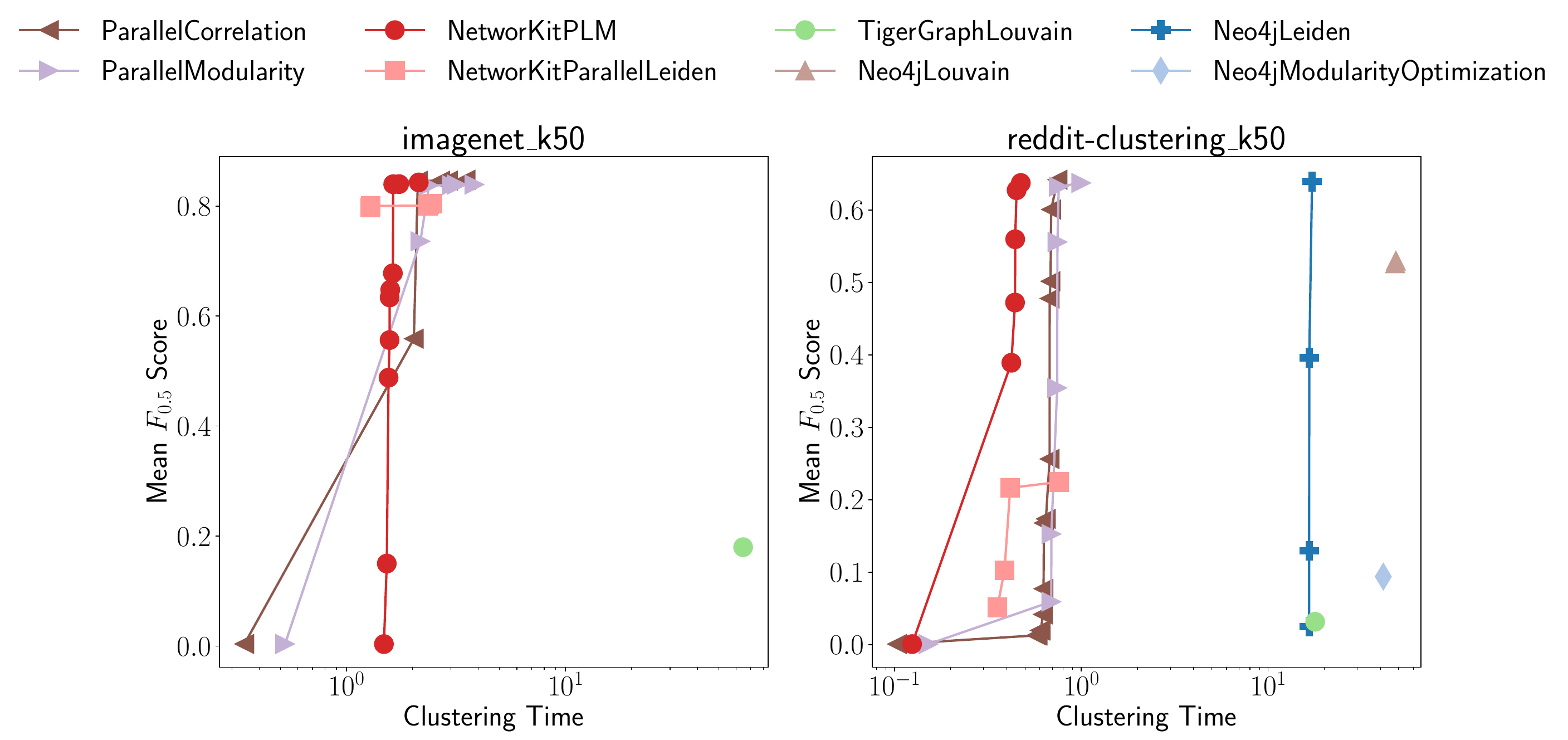}
    \vspace{-2.5em}
    \caption{The Pareto frontiers for the unweighted SNAP graphs, using different modularity implementations. }
    \label{fig:modularity_subset}
\end{figure}

In \Cref{fig:modularity_subset}, we show the precision and recall comparison of different modularity implementations on a subset of datasets. The results on all datasets can be found in 
\iffull
\Cref{fig:snap_modularity} and \Cref{fig:large_weight_appendix} in \Cref{sec:appendix-mod}. 
\fi
SnapCNM is only able to finish on \datasetname{MNIST} so it is only shown on a single subplot.

We see that \parc{}'s correlation clustering and modularity clustering implementations are competitive with NetworKit's PLM and NetworKit's ParallelLeiden implementations. Other modularity objective-based methods are much slower and do not achieve higher quality. Methods that do not have a resolution parameter have unstable clustering quality (Neo4jModularityOptimization, Neo4jLouvain, and TigerGraphLouvain).

Compared to NetworKitPLM, \parc{} (ParallelCorrelation and ParallelModularity) is slightly slower on the weighted graphs, but is faster and has higher quality on the unweighted graphs. Compared to NetworKitParallelLeiden, \parc{} has higher quality on most datasets, and has similar running time.

Finally, we compare the modularity objective obtained by different modularity methods. In \Cref{tab:modularity_scores}, we show the modularity scores with $\gamma=1$ for \datasetname{LJ}, using different modularity methods.
\iffull
In \Cref{fig:modularity_scores_all},
\fi
 we show the scores for all unweighted graphs. We observe that PCBS's modularity clustering, Neo4jLouvain, NetworKitPLM, and Neo4jLeiden get similar scores, and are the highest ones.
Neo4jModularity, and NetworKitLeiden get similar scores, and are slightly lower than the methods above.
TigerGraphLouvain gets the lowest score.

\begin{table}[t]
\small
    \centering
    \caption{The modularity scores with $\gamma=1$ for \datasetname{LJ}, using different modularity methods.}
    \vspace{-1em}
    \begin{tabular}{lr}
    \toprule
     Clusterer & Modularity Objective \\
    \midrule
    PCBS Modularity & 0.755035 \\
    Neo4j Louvain & 0.754733 \\
    NetworKit PLM & 0.753670 \\
    Neo4j Leiden & 0.743583 \\
    NetworKit Leiden & 0.662478 \\
    Neo4j Modularity & 0.654594 \\
    TigerGraph Louvain & 0.022660 \\
    \bottomrule
    \end{tabular}
    \label{tab:modularity_scores}
\end{table}

%% file: datasets.tex
\begin{table}[t]
\small
\centering
\vspace{1em}
\caption{Unweighted graph datasets from SNAP~\cite{leskovec2016snap}. }
\vspace{-1em}
\begin{tabular}{|l|r|r|}
\hline
\textbf{Dataset} & \textbf{Num. Vertices} & \textbf{Num. Edges} \\
\hline
Amazon (AM) & 334,863 & 925,872  \\
\hline
com-DBLP (DB) & 425,957 & 2,099,732  \\
\hline
YouTube-Sym (YT) & 1,138,499 & 5,980,886  \\
\hline
LiveJournal (LJ) & 4,847,571 & 85,702,474  \\
\hline
com-Orkut (OK) & 3,072,627 & 234,370,166  \\
\hline
Friendster (FS) & 65,608,366 & 3,612,134,270  \\
\hline
\end{tabular}
\label{table:graph_datasets}
\end{table}

\begin{table}[t]
\small
\centering
\vspace{1em}
\caption{Embedding datasets from \citet{yu2023pecann}. "Num. Clusters" is the number of ground truth clusters. }
\vspace{-1em}
\label{table:weighted_datasets}
\begin{tabular}{|l|r|r|r|}
\hline
\textbf{Dataset} & \textbf{Num. Vertices} & \textbf{Num. Clusters}  \\
\hline
MNIST & 70,000 & 10  \\
\hline
StackExchange & 373,850 & 121 \\
\hline
Reddit & 420,464 & 50 \\
\hline
ImageNet & 1,281,167 & 1000  \\
\hline
\end{tabular}

\end{table}

\begin{table}[t]
\centering
\caption{Synthetic RMAT graphs.}
\vspace{-1em}
\label{table:rmat_datasets}
\small
\begin{tabular}{|l|r|r|r|}
\hline
\textbf{Dataset} & \textbf{Num. Vertices} & \textbf{Num. Edges} & \textbf{Avg. Degree} \\
\hline
RMAT-1 & $2^{15}$ & 31,605,326 & 964.5 \\
\hline
RMAT-2 & $2^{21}$ & 396,360,032 & 189.0\\
\hline
\end{tabular}
\end{table}

%% file: table_platforms.tex
\rowcolors{2}{gray!25}{white}
\begin{table*}[t]
\small
    \centering
\caption{\small Undirected graph clustering algorithms implemented in \parc{} and popular graph libraries and databases. The starred (*) implementations are sequential. The rows are sorted by the number of ticks. We exclude clustering algorithms that do not consider graph edges (e.g., $k$-means clustering). Modularity Clustering includes both Louvain and Leiden implementations. 
Hop Preference \& Node Preference (HANP)~\cite{leung2009towards} and Conductance Minimization~\cite{soman2011fast} are variants of the Label Propagation algorithm. Neo4j implements a parallel version of the sequential maximum $k$-cut algorithm~\cite{festa2002randomized}.
    Memgraph, NebulaGraph, and Oracle Graph are graph databases  described in \Cref{sec:related}.
    }
\vspace{-1em}
\begin{tabular}{c|c|c|c|c|c|c|c|c}
\hline
\multicolumn{1}{c|}{} & \multicolumn{3}{c|}{Graph Libraries} & \multicolumn{5}{c}{Graph Databases} \\
\hline
& \textbf{Ours} & \textbf{NetworKit} & \textbf{SNAP} & \textbf{Neo4j} & \textbf{NebulaGraph} & \textbf{Oracle Graph} & \textbf{TigerGraph} & \textbf{Memgraph} \\
\hline
Modularity Clustering & \checkmark & \checkmark & \checkmark* & \checkmark & \checkmark & \checkmark & \checkmark & \checkmark \\
Connectivity & \checkmark & \checkmark & \checkmark* & \checkmark & \checkmark & \checkmark & \checkmark & \checkmark \\
KCore & \checkmark & \checkmark & \checkmark* & \checkmark & \checkmark & \checkmark & \checkmark & \\
Label Propagation & \checkmark & \checkmark & & \checkmark & \checkmark & \checkmark & \checkmark & \\
SLPA & \checkmark & & & \checkmark & & & \checkmark & \\
Infomap & & & \checkmark* & & \checkmark & \checkmark & & \\
SCAN & \checkmark & & & & & & & \\
TECTONIC & \checkmark & & & & & & & \\
Unweighted LDD & \checkmark & & & & & & & \\
HAC & \checkmark & & & & & & & \\
Affinity Clustering & \checkmark & & & & & & & \\
Correlation Clustering & \checkmark & & & & & & & \\
Approximate Maximum $k$-cut & & & & \checkmark & & & & \\
Conductance Minimization & & & & & & \checkmark & & \\
HANP & & & & & \checkmark & & & \\
\hline
\end{tabular}
\label{tab:all_platforms}
\end{table*}

%% file: primer.tex
\section{Related Work}\label{sec:related}

\revised{\myparagraph{Clustering using Graph Processing Libraries} Libraries for graphs, such as NetworkX~\cite{hagberg2008exploring}, SNAP~\cite{leskovec2016snap}, GBBS~\cite{gbbs}, GraphMineSuite~\cite{GraphMineSuite}, and NetworKit~\cite{staudt2016networkit}, are software packages that provide data structures and (parallel) algorithms for representing and operating on graphs. 
These libraries primarily focus on in-memory computation and analysis of graph data, offering flexibility to build custom graph processing applications and algorithms.
PCBS is a clustering library built on top of GBBS, an existing graph processing library; our results show that in general PCBS implementations are competitive or faster than the fastest existing library implementations of many graph clustering algorithms (see Section~\ref{sec:exp}).

}

\revised{
\myparagraph{Clustering using Graph Databases}
Graph databases are specialized database management systems designed explicitly for storing and querying highly interconnected data represented as nodes, edges, and properties.  
Graph databases support declarative graph query languages like Cypher, Gremlin, and SPARQL, and are optimized for persisting and retrieving graph data. 
Examples of popular graph databases include Neo4j~\cite{neo4j,Dominguez-Sal10}, TigerGraph~\cite{tigergraph}, Amazon Neptune~\cite{neptune},  Memgraph~\cite{memgraph}, OrientDB~\cite{orientdb}, ArangoDB~\cite{arangodb}, and NebulaGraph~\cite{nebulagraph}. \citet{besta2019demystifying} provide a comprehensive survey of graph databases.
Compared with clustering implementations in graph databases, PCBS implementations are significantly faster, likely due to a combination of using efficient parallel data structures and algorithms, and not paying for any storage and data structural overheads incurred by existing graph database formats.
}

\revised{
\myparagraph{Prior Studies on Graph Clustering}
Due to the importance of the clustering problem and the abundance of different algorithms, there are many comparative studies on clustering algorithms.
\citet{yang2012defining} compare clustering algorithms on social, collaboration, and information networks. Unlike our work, their work focuses primarily on community scoring functions and does not compare the running times of the algorithms. 
\citet{fortunato2010community} survey clustering methods in graphs and compared many algorithms; unlike the present work, they do not present experiments for parallel methods. 
There are also many works that evaluate algorithms on artificial networks~\cite{shi2020comparison, yang2016comparative, orman2011qualitative, orman2012comparative, Maekawa2019, park2023identifying}.  
Lastly, there are also several graph clustering benchmarks for overlapping clusters and heterogeneous graphs
~\cite{xie2013overlapping, knudsen2022grab, lancichinetti2009detecting}. 
}

\revised{The aforementioned works usually only compare the quality or running time of the algorithms, but not both simultaneously unlike this work.
For the works that compare both, they 
do not experiment with parallel algorithms or study scalability with increasing core (or machine) count.}
\citet{Bader2014benchmarking}'s graph clustering challenge has an objective function, where both quality and speed contributed to the final scores, but their quality measures do not compare with ground truth clusters. Moreover, they only released graph datasets but did not provide comprehensive experimental results.
\revised{As far as we know, our work is the first to study parallel graph clustering algorithms both in terms of quality and running time.}

%% file: appendix-scalability.tex
\section{Runtime and Scalability of all algorithms}\label{sec:runtime-appendix}

\begin{figure*}[t]
    \centering
    \includegraphics[width=0.49\textwidth]{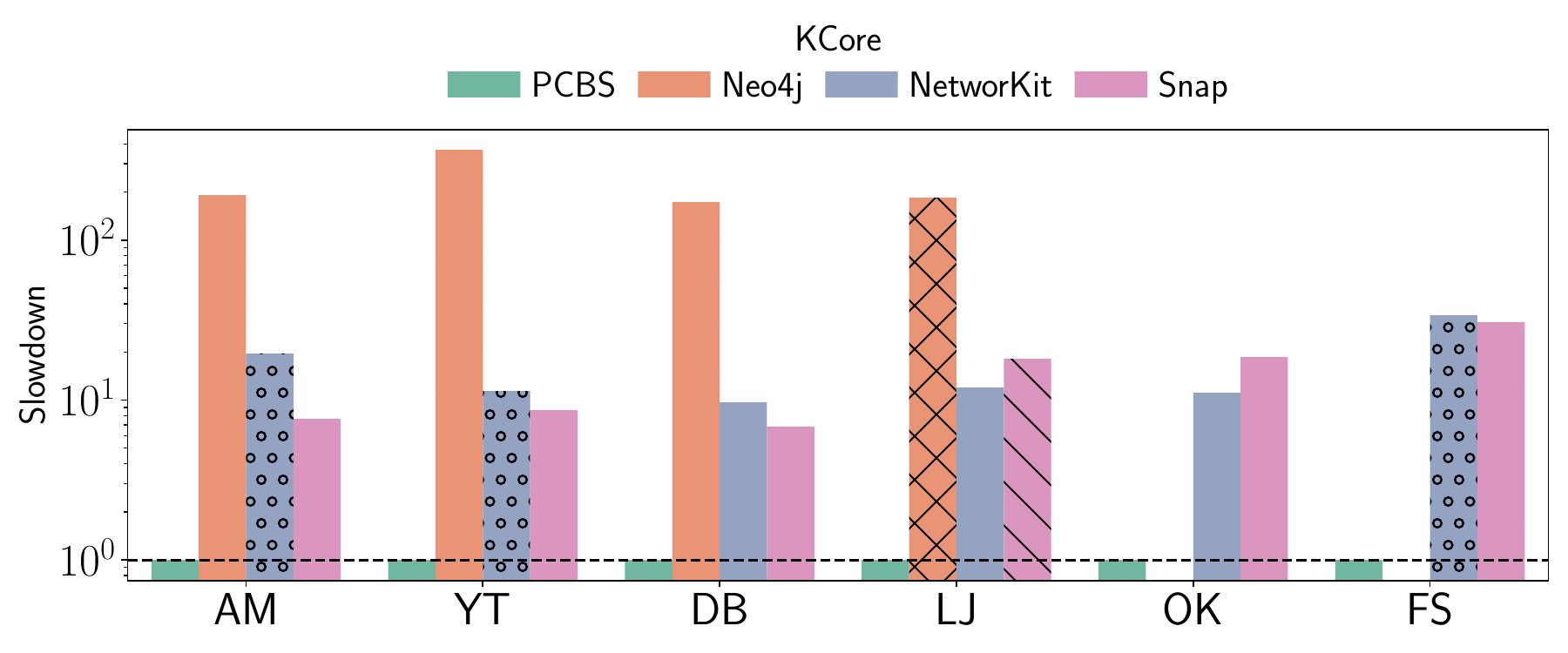}
    \includegraphics[width=0.49\textwidth]{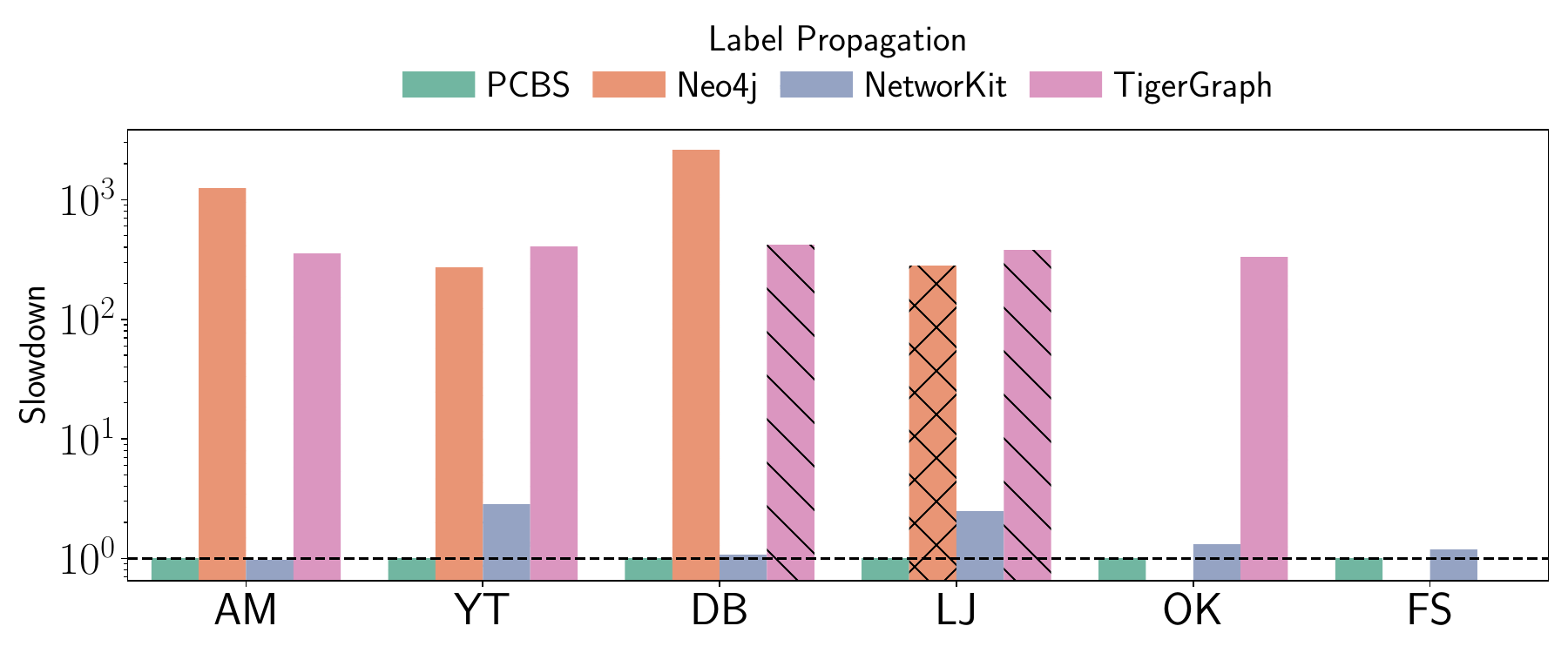}
    \includegraphics[width=0.49\textwidth]{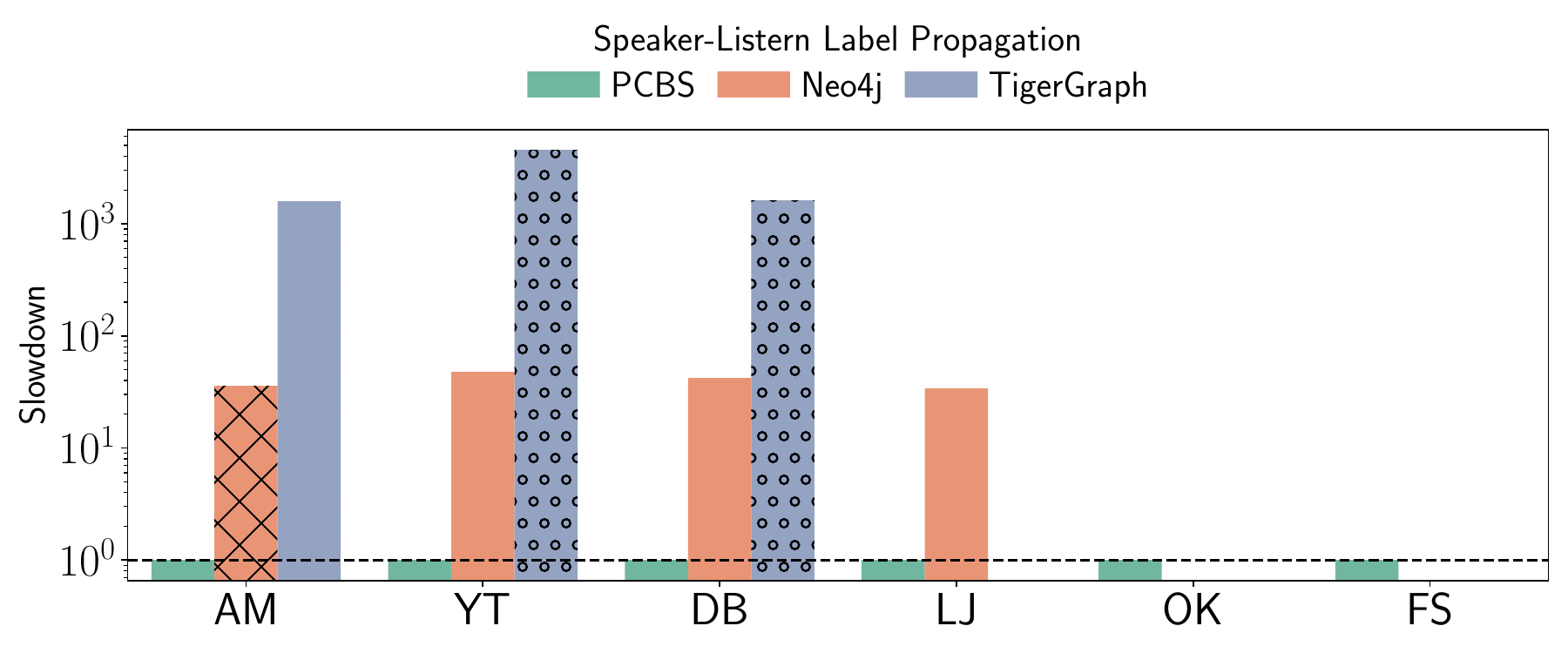}
    \caption{Slowdown of methods on the unweighted graphs. }
    \label{fig:slowdown-all}
\end{figure*}

In \Cref{fig:slowdown-all}, we show the running time comparisons of different implementations of LP and SLPA on the unweighted graphs. \parc{}, NetworKit, and TigerGraph use all 60 threads. SNAP is sequential. Neo4j only uses 4 cores due to community edition limitations. 
We see that \parc{}'s implementations are the fastest.

\begin{figure*}[t]
    \centering
    \includegraphics[width = \textwidth]{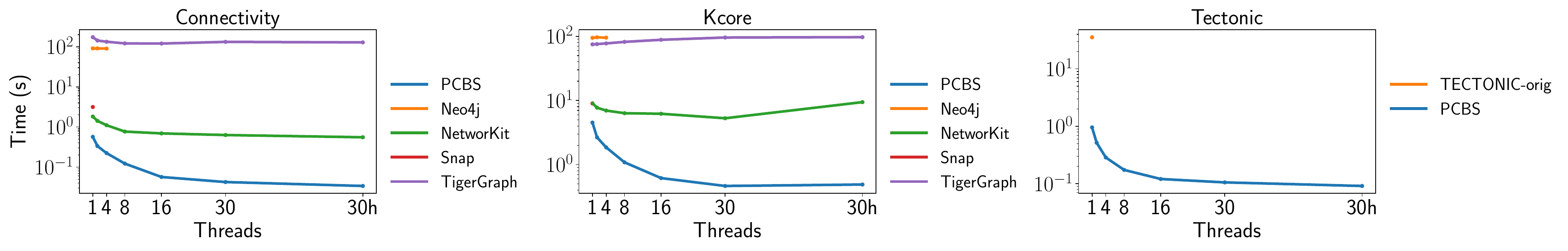}
    \includegraphics[width = \textwidth]{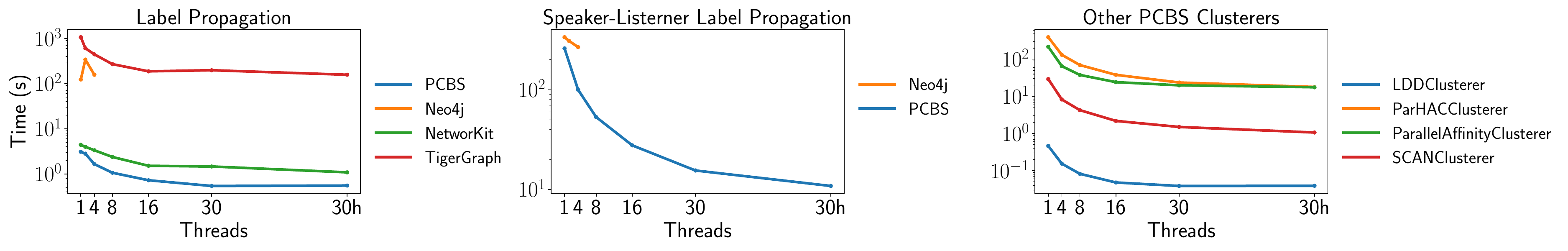}
    \caption{Running time of the clustering algorithms. }
    \label{fig:scalability}
\end{figure*}

In \Cref{fig:scalability}, we show the running time of different implementations using different numbers of threads. For all plots except Tectonic, \datasetname{livejournal} graph is used. For Tectonic, \datasetname{youtube} graph is used because the original TECTONIC implementation fails to run on \datasetname{livejournal}. 

We see that for all methods experimented in \Cref{fig:scalability}, \parc{}'s implementations are the fastest on all threads and they have good scalability with respect to the number of threads. 

%% file: appendix-k.tex
\section{Using different $k$ for $k$-nearest neighbor graphs}\label{sec:compare_k}
\begin{figure*}
    \centering
    \includegraphics[width=\textwidth]{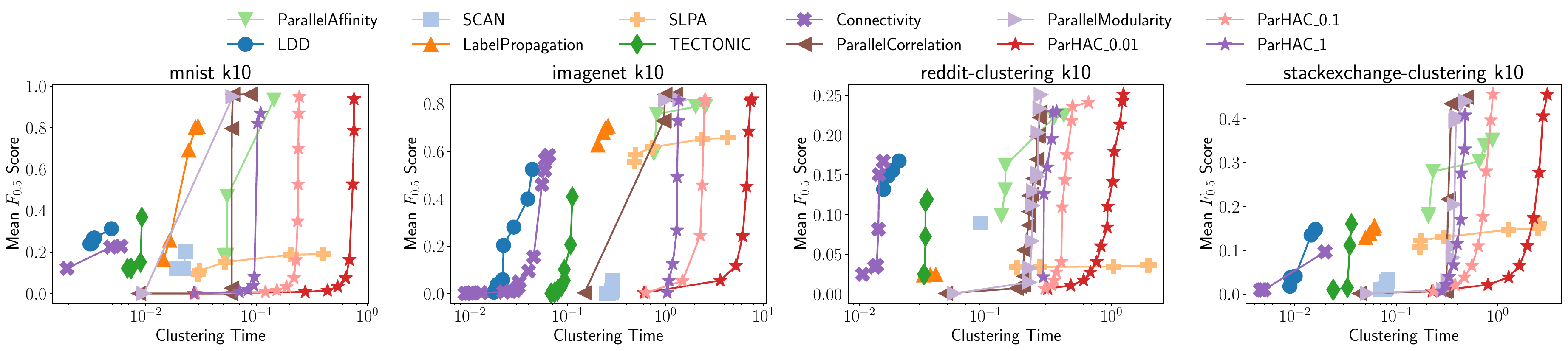}
    \includegraphics[width=\textwidth]{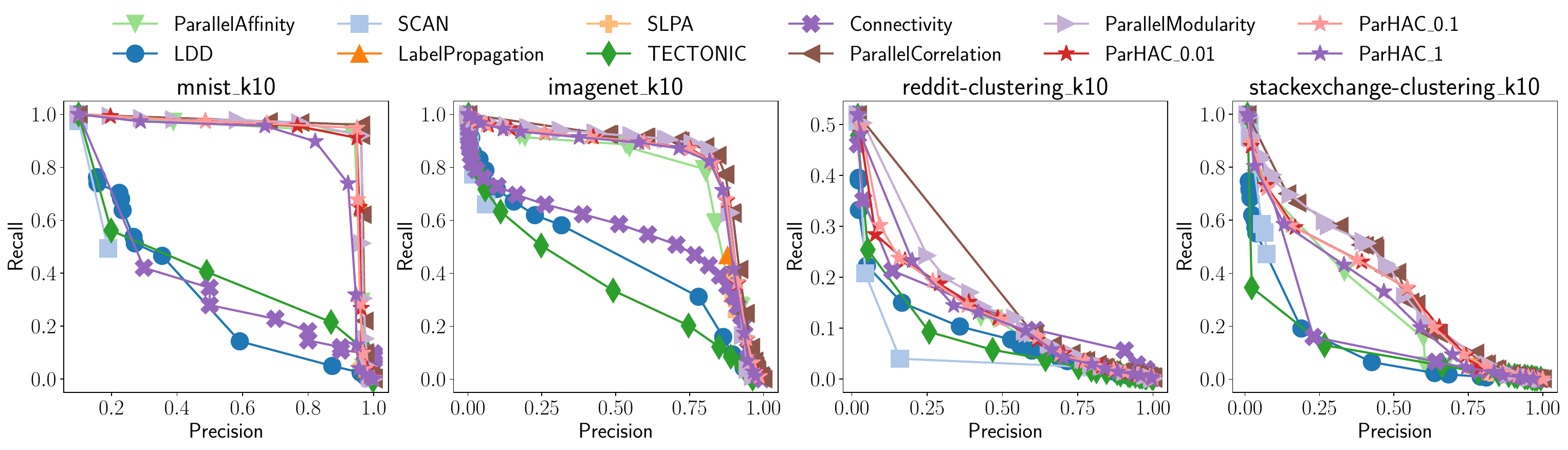}
    \caption{The Pareto frontier graphs for the weighted \knn graphs ($k=10$), using \parc{} methods.}
    \label{fig:k_10}
\end{figure*}

\begin{figure*}
    \centering
    \includegraphics[width=\textwidth]{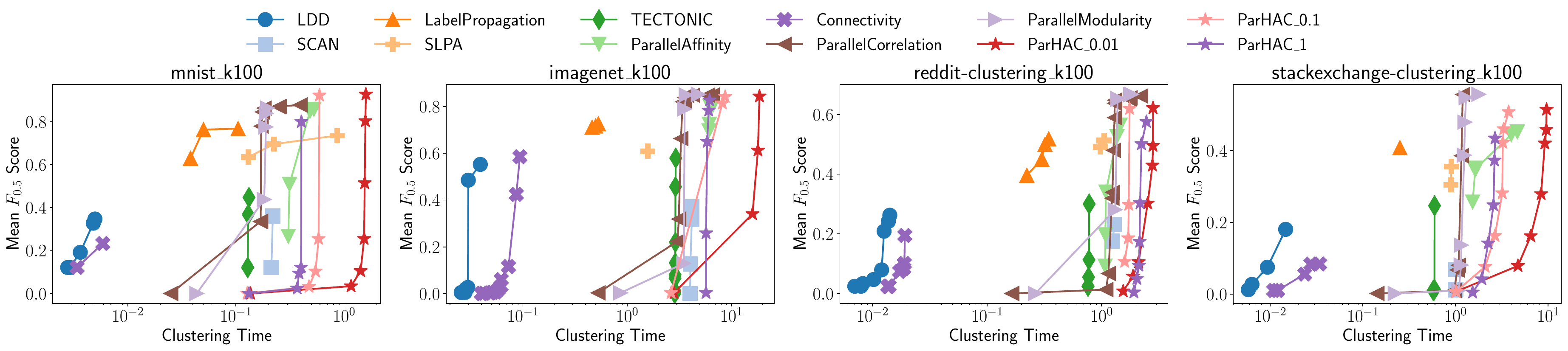}
    \includegraphics[width=\textwidth]{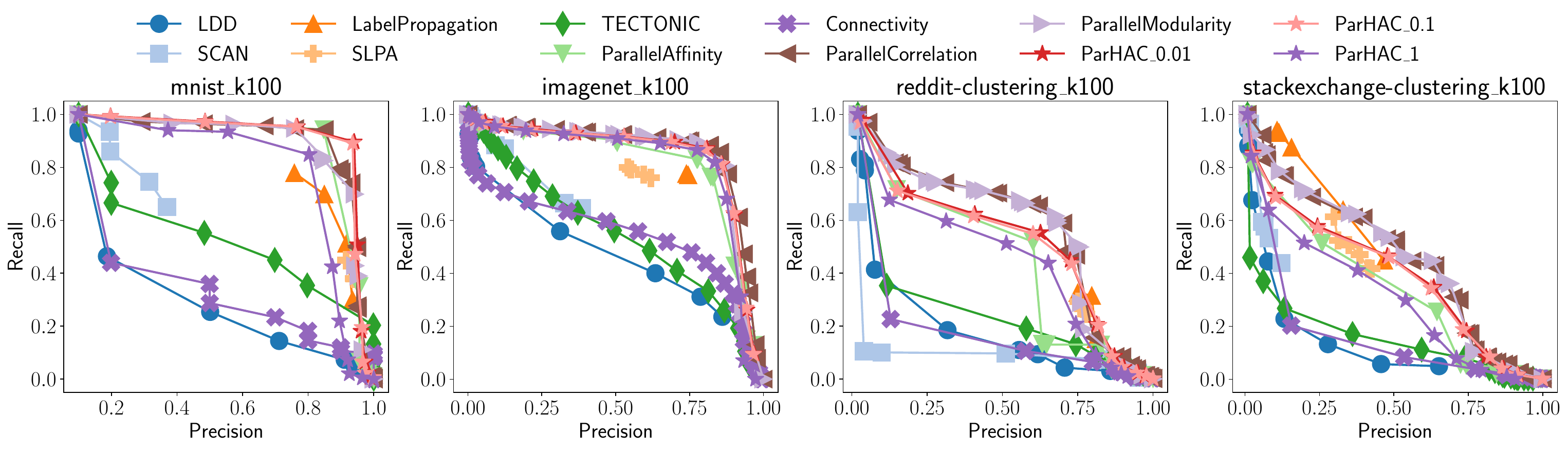}
    \caption{The Pareto frontier graphs for the weighted \knn graphs ($k=100$), using \parc{} methods.}
    \label{fig:k_100}
\end{figure*}

In \Cref{fig:k_10,fig:k_100} we present the result on the \knn graphs when $k=10$ and $k=100$. 

We observe that for the weighted graphs with a small number of ground truth clusters, increasing $k$ from 10 to 50 improves the clustering quality of \datasetname{reddit} and \datasetname{stackexchange}. Increasing $k$ more from 50 to 100 doesn't significantly improve the clustering quality of any of the four data sets. Moreover, for all three $k$ values (10, 50, and 100) the comparisons between the clustering algorithms generally stay the same.

%% file: appendix-datasets.tex
\section{More Details on Dataset}\label{appendix-dataset}
The datasets in \Cref{table:weighted_datasets} are obtained from \citet{yu2023pecann}.
\datasetname{MNIST}~\cite{deng2012mnist} is a standard machine learning dataset that consists of $28 \times 28$ dimensional images of grayscale digits between $0$ and $9$. The $i^\text{th}$ cluster corresponds to all occurences of digit $i$.
\datasetname{ImageNet}~\cite{imagenet} is a standard image classification benchmark with more than one million images, each of size $224 \times 224 \times 3$. The images are from $1000$ classes of everyday objects. Unlike for MNIST, we do not cluster the raw \datasetname{ImageNet} images, but instead first pass each image through  ConvNet~\cite{liu2022convnet} to get an embedding. Each ground truth cluster contains the embeddings corresponding to a single image class from the original \datasetname{ImageNet} dataset. \datasetname{Reddit} and \datasetname{StackExchange} are text embedding datasets studied in the recent Massive Text Embedding Benchmark  (MTEB) work~\cite{muennighoff2022mteb}. We restrict our attention to embeddings from the best model on the current MTEB leaderboard, GTE-large~\cite{li2023towards}.

The datasets in \Cref{table:weighted_datasets} use ParlayANN's Vamana graph method~\cite{manohar2024parlayann} to compute the approximate $k$-nearest neighbors.

\begin{figure*}
    \centering
    \includegraphics[width=\textwidth]{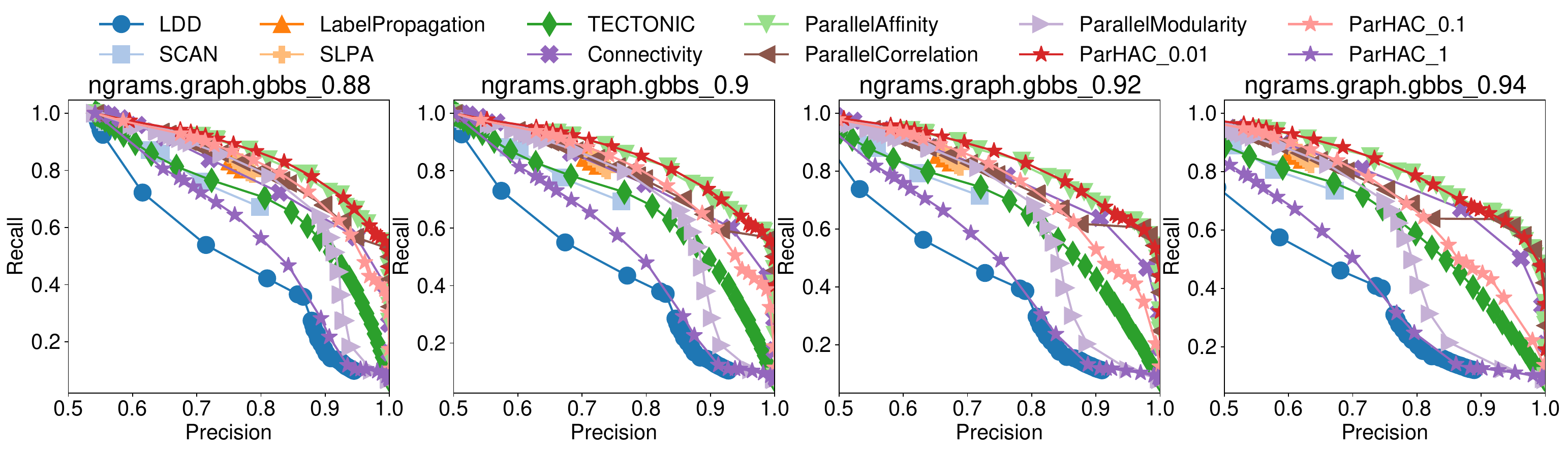}
    \caption{\revised{Precision and Recall pareto frontier with respect to different thresholds.The threshold is used to generate positive and negatives pairs in the ground truth.}}
    \label{fig:pr_ngrams}
\end{figure*}

\revised{
\subsection{NGrams Similarity Graph Dataset}

In this subsection, we describe how we generated our NGrams similarity graph.

\begin{enumerate}
    \item \textbf{Data Source:} We start with a text dataset of ngrams~\cite{goldberg2013dataset}, and select all ngrams that occur at least 120 times in the verbargs part of the dataset. This results in a corpus of 1,274,126 short texts.

    \item \textbf{Embedding:} Each ngram is then embedded using the text-embedding-gecko@003 model, specifying \texttt{CLUSTERING} as the \texttt{task\_type}.
    This transforms the text data into 768-dimensional vector representations that capture semantic meaning.
    The similarity between two embeddings is obtained by computing their dot product.

    \item \textbf{$k$-Nearest Neighbor Graph Construction:} We compute an exact 50-nearest neighbor graph and make it undirected. 

    \item \textbf{Label Generation} Finally, we build a set of labels that we later use to evaluate clustering quality, by performing the following steps:
    \begin{itemize}
        \item Define similarity buckets: $[0.76, 0.77)$, $[0.77, 0.78)$, ..., $[0.99, 1)$
        \item Sample $100\,000$ embeddings uniformly at random.
        \item For each sampled embedding $x$ and each similarity bucket $[s, s+0.01)$, among all embeddings whose similarity to $x$ belongs to $[s, s+1)$ sample one embedding $y$ uniformly at random (if one exists). We represent this sample with a tuple $(x, y, x \cdot y)$, where $x \cdot y$ is the similarity of $x$ and $y$.
        \item From all the tuples created this way we sample $4\,000$ in each similarity bucket uniformly at random. This yields $96\,000$ tuples in total. This approach ensures diversity in both the points selected and the range of similarities represented.
        \item  The 96\,000 tuples are converted into labels based on a similarity threshold $t$. Pairs with similarity larger than the threshold are labeled as belonging to the same cluster, while those below this threshold are labeled as belonging to different clusters.
    We used threshold = 0.92 when calculating the precision and recall in \Cref{sec:exp}. In \Cref{fig:pr_ngrams}, we also show the results for thresholds in [0.88, 0.90, 0.92, 0.94], we see that the results are similar. 
    \end{itemize}

    We note that using simpler sampling approaches leads to a sample that we consider less representative.
    For example, if we sampled pairs of points uniformly at random, we would end up with very few pairs of high similarity.
    A simple fix to this approach is to uniformly sample pairs of points whose similarity falls into each similarity bucket.
    However, this may lead to some points occurring in the sample much more often than others.
    In our sample, each input ngram occurs in at most $9$ labels.
\end{enumerate}

When computing the precision and recall, given a threshold $t$, for each tuple $(x, y, w)$ the pair $(x, y)$ is called positive if $w > t$ and negative otherwise.

Then, for a positive pair $(x, y)$, if  $x$ and $y$  are in the same cluster, then we have a true positive; otherwise we have a false negative.
For a negative pair $(x, y)$, if $x$ and $y$ are in different clusters, then we have a true negative; otherwise we have a false positive.
Finally, precision and recall are computed the usual way. 

We note that the dataset we release consists of three files:
\begin{itemize}
\item The ngrams and their corresponding embeddings.
\item Edges of the symmetrized 50-NN graph.
\item The $96\,000$ sampled tuples, which can be used to obtain labels.
\end{itemize}
}

\begin{table}[]
\caption{\revised{Area under curve for precision $\geq 0.5$ on the NGram graph with $k=50$. 
}}
\vspace{-1em}
    \centering
    \small
\begin{tabular}{lr}
\toprule
ParHAC-0.01  &   \textbf{0.83} \\
Affinity     &              0.82 \\
Correlation  &              0.77 \\
Connectivity &              0.77 \\
ParHAC-0.1   &              0.74 \\
TECTONIC     &              0.64 \\
Modularity   &              0.63 \\
SLPA         &              0.58 \\
LP           &              0.57 \\
SCAN         &              0.56 \\
ParHAC-1     &              0.46 \\
LDD          &              0.38 \\
\bottomrule
\end{tabular}
    \label{tab:ngram}
\end{table}

\section{More Details on Experiments}
We used Neo4j community version 5.8.0 with graph data science library version 2.4.3 and TigerGraph version 3.9.2.

For $k$-core clustering, all baseline implementations only return the $k$-core decomposition and do not return a clustering. We implement a $k$-core clustering for NetworKit using its parallel connected components. For Neo4j and TigerGraph, we report only the $k$-core decomposition time.

%% file: appendix-exp.tex
\section{Experiments on UCI datasets}\label{sec:uci}

\begin{figure*}
    \centering
    \includegraphics[width=\textwidth]{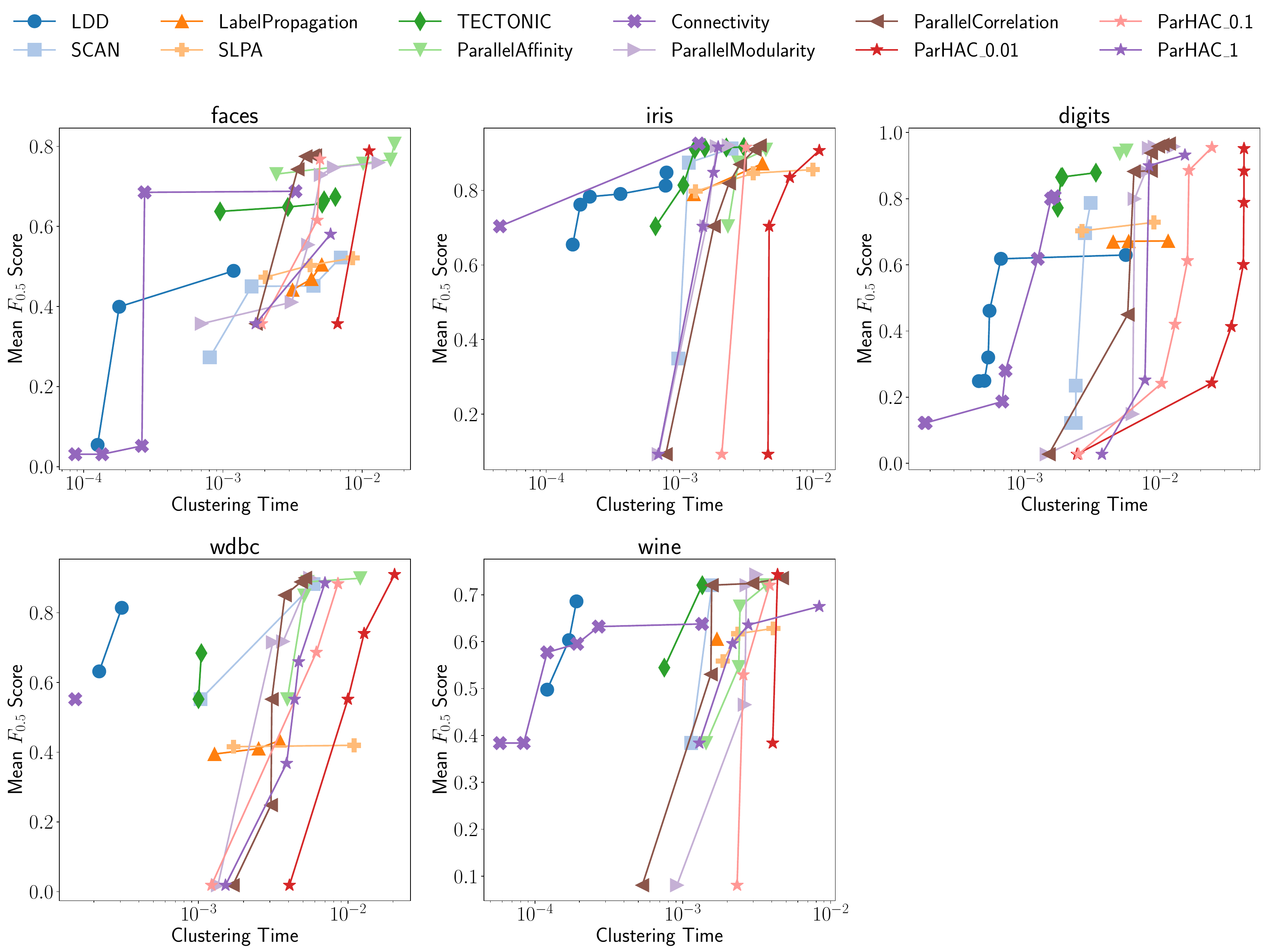}
    \caption{The Pareto frontier of the $F_{0.5}$ and runtime for the weighted UCI \knn graphs $k=10$.}
    \label{fig:uci_f1}
\end{figure*}

In \Cref{fig:uci,fig:uci_f1}, we present the results on five small UCI~\cite{asuncion2007uci} $k$-nearest neighbor graphs. A summary of the datasets is in \Cref{table:uci_datasets}. We see that on these small data sets, the quality and runtime difference is not as significant as in the larger weighted graph as shown in \Cref{sec:weighted}.

\begin{table}[t]
\centering
\caption{Description of the weighted UCI datasets. $k=10$ is used to generate the $k$-nearest neighbor graphs.}
\begin{tabular}{|l|l|l|l|}
\hline
\textbf{UCI Dataset} & \textbf{Num. Vertices} & \textbf{Num. Cluster}\\
\hline
faces & 400 & 40 \\
\hline
iris & 150 & 3 \\
\hline
wine & 178 &  3 \\
\hline
wdbc & 569 & 2  \\
\hline
digits & 1797 & 10 \\
\hline
\end{tabular}
\label{table:uci_datasets}
\end{table}

\begin{table}[t]
\centering
\caption{The AUC scores for the weighted UCI \knn graphs $k=10$. }
\begin{tabular}{l|lccccc}
\toprule
 & faces & iris & digits & wdbc & wine & Mean \\
\midrule
Correlation & 0.58 & 0.93 & \textbf{ 0.94 } & 0.85 & \textbf{ 0.54 } & \textbf{ 0.77 } \\
Modularity & 0.58 & \textbf{ 0.94 } & 0.93 & 0.86 & 0.52 & 0.77 \\
ParHac-0.01 & 0.60 & 0.92 & 0.92 & \textbf{ 0.87 } & 0.48 & 0.76 \\
Affinity & \textbf{ 0.63 } & 0.92 & 0.91 & 0.82 & 0.48 & 0.75 \\
ParHac-0.1 & 0.58 & 0.86 & 0.92 & 0.81 & 0.50 & 0.73 \\
ParHac-1 & 0.32 & 0.90 & 0.90 & 0.80 & 0.44 & 0.67 \\
Tectonic & 0.53 & 0.94 & 0.87 & 0.48 & 0.49 & 0.66 \\
Scan & 0.35 & 0.90 & 0.65 & 0.80 & 0.49 & 0.64 \\
Connectivity & 0.53 & 0.86 & 0.71 & 0.50 & 0.43 & 0.61 \\
LDD & 0.24 & 0.59 & 0.44 & 0.50 & 0.48 & 0.45 \\
SLPA & 0.35 & 0.57 & 0.39 & 0.14 & 0.27 & 0.34 \\
LP & 0.36 & 0.59 & 0.34 & 0.14 & 0.25 & 0.34 \\
\bottomrule
\end{tabular}
\label{tab:uci_auc}
\end{table}

\begin{figure*}
    \centering
    \includegraphics[width=\textwidth]{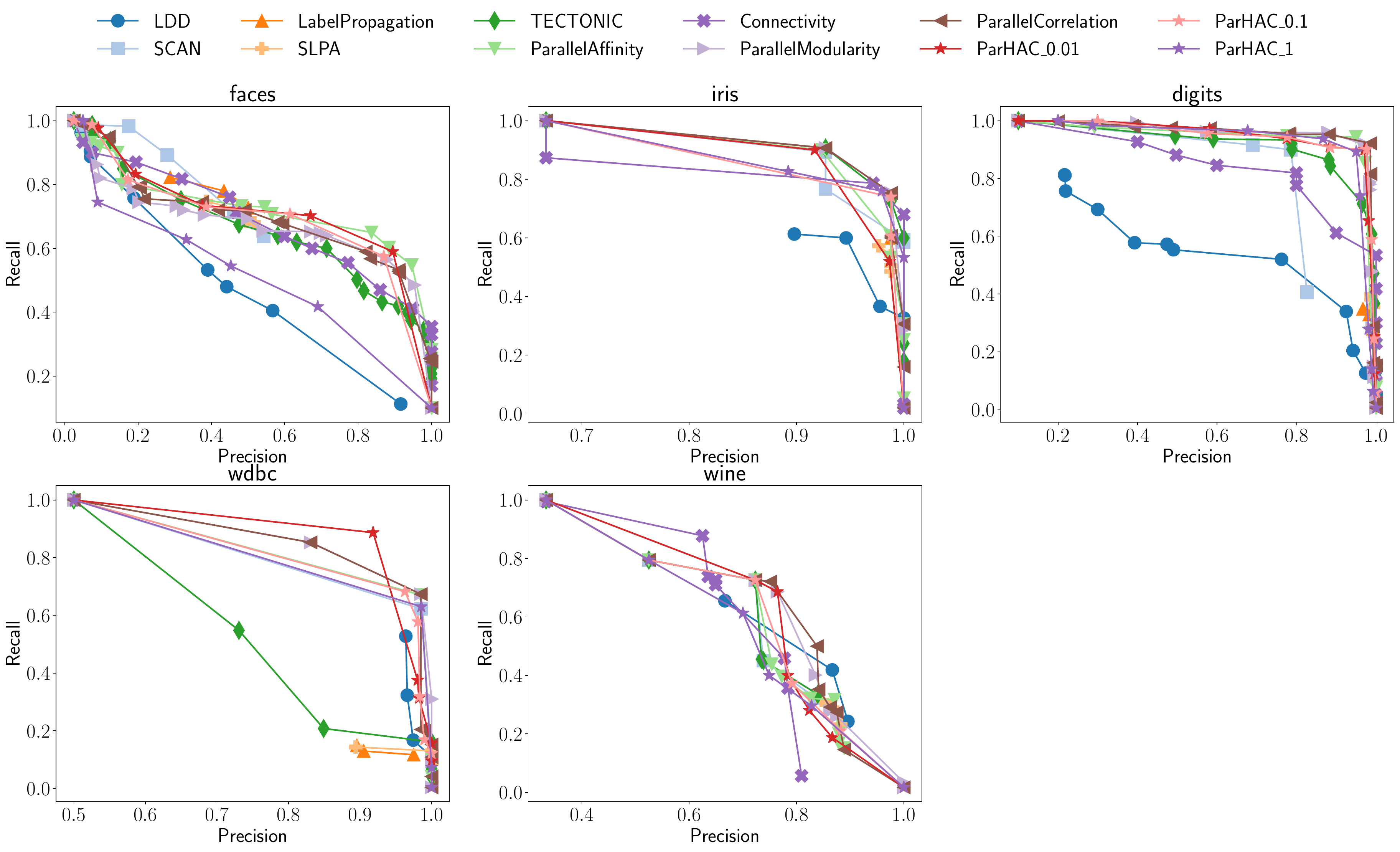}
    \caption{The Pareto frontier of the precision and recall for the weighted UCI \knn graphs $k=10$. }
    \label{fig:uci}
\end{figure*}

\begin{figure*}
    \centering
    \includegraphics[width=\textwidth]{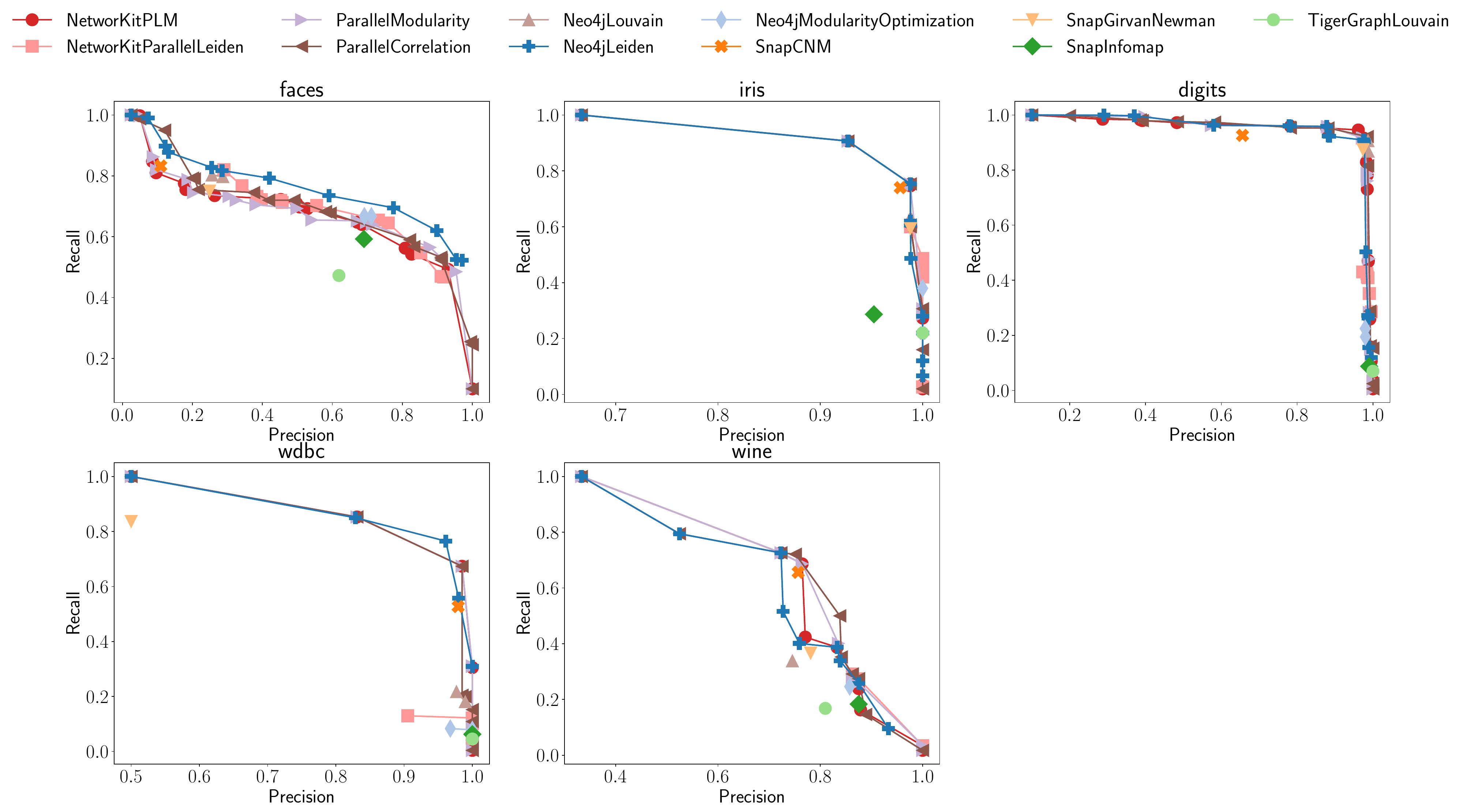}
    \caption{The Pareto frontier of the precision and recall for the weighted UCI \knn graphs ($k=10$), using different modularity methods.}
    \label{fig:uci}
\end{figure*}

\begin{figure*}[t]
    \centering
    \includegraphics[width=\textwidth]{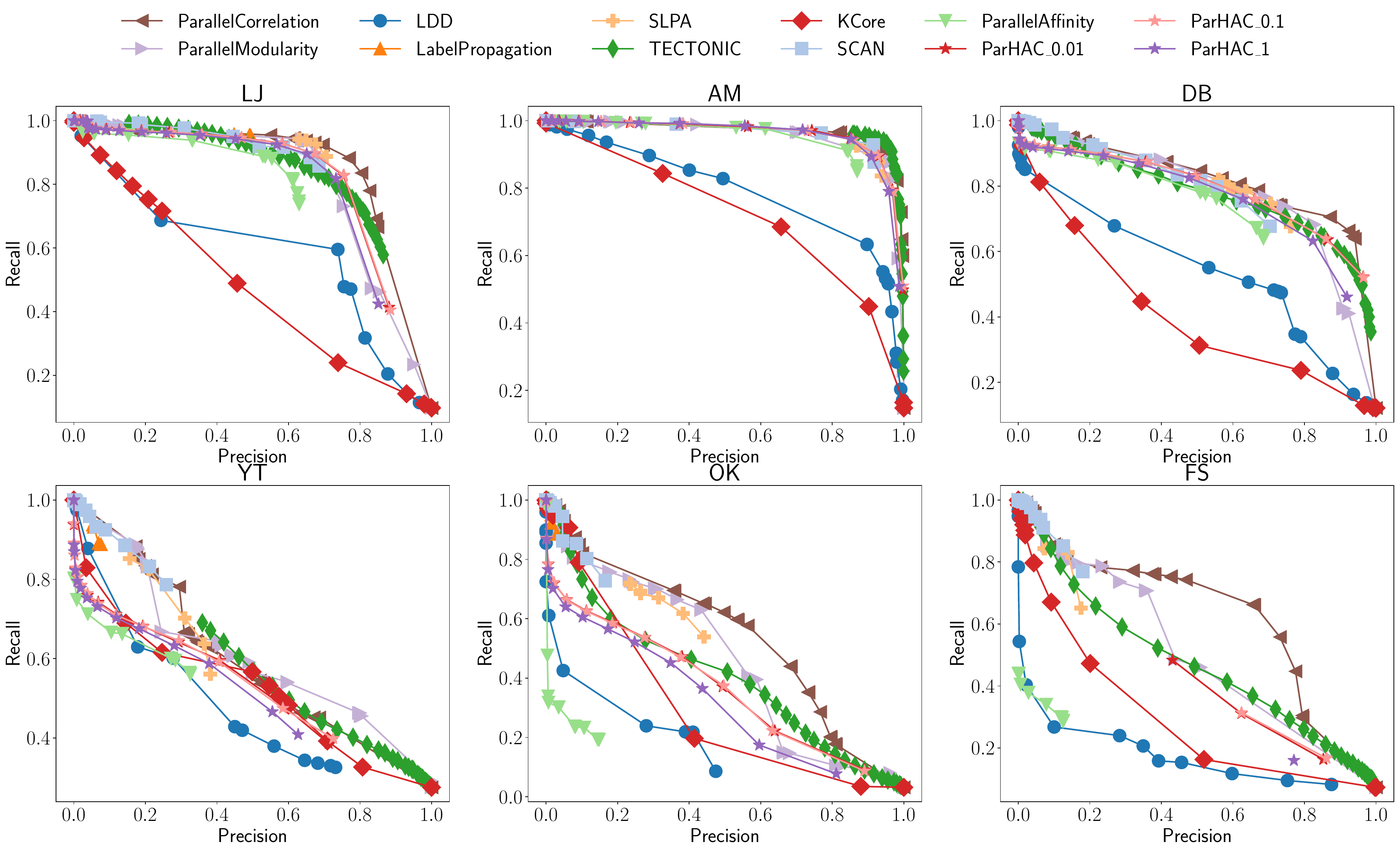}
    \caption{The Pareto fronriter of the precision and recall of the unweighted SNAP graphs.}
    \label{fig:pr_snap}
\end{figure*}

\begin{figure*}[t]
    \centering
    \includegraphics[width=\textwidth]{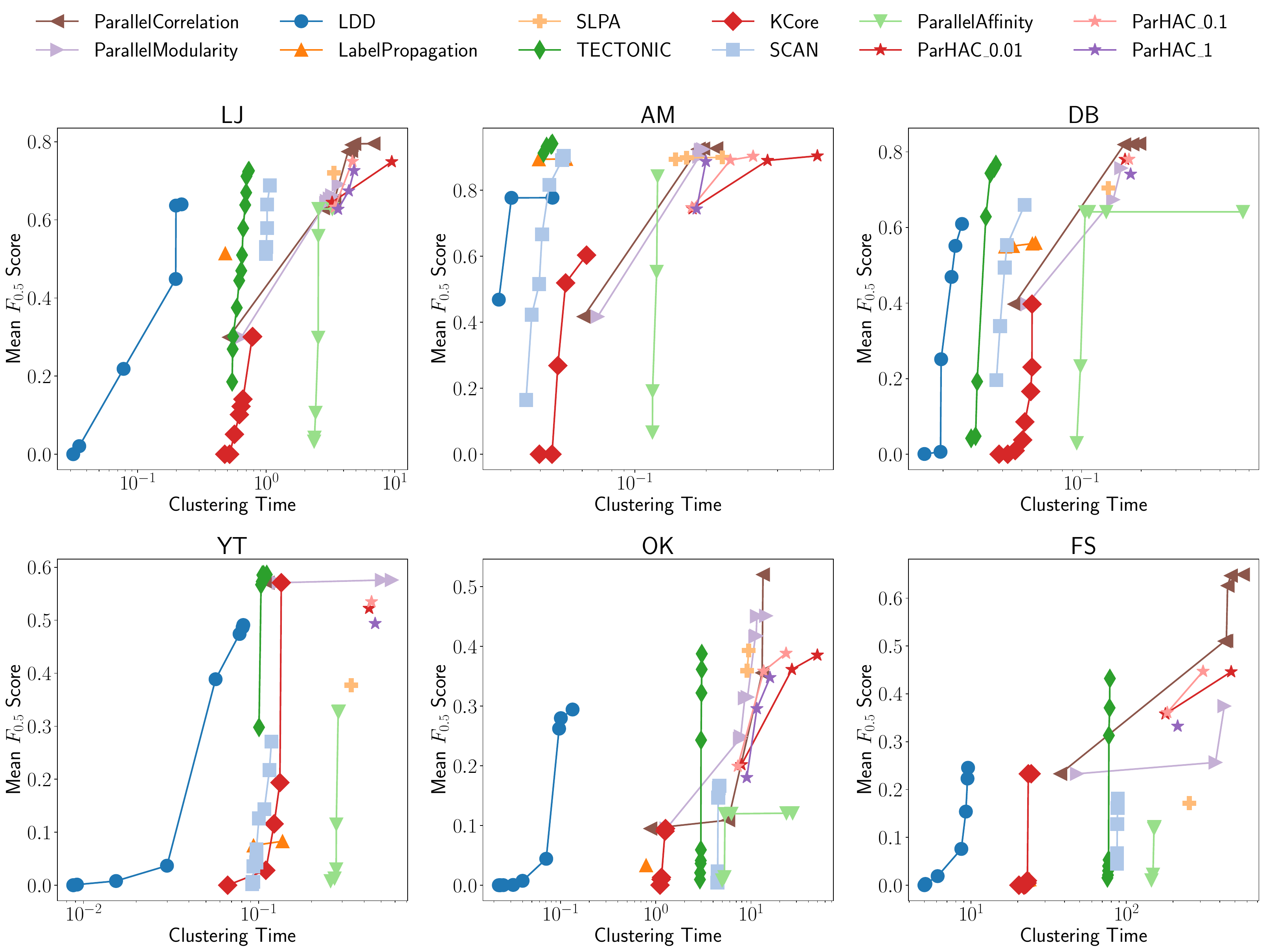}
    \caption{The Pareto frontier of the $F_{0.5}$ and runtime graph for the unweighted SNAP graphs. }
    \label{fig:time_f1_snap}
\end{figure*}

\begin{figure*}
    \centering
    \includegraphics[width=\textwidth]{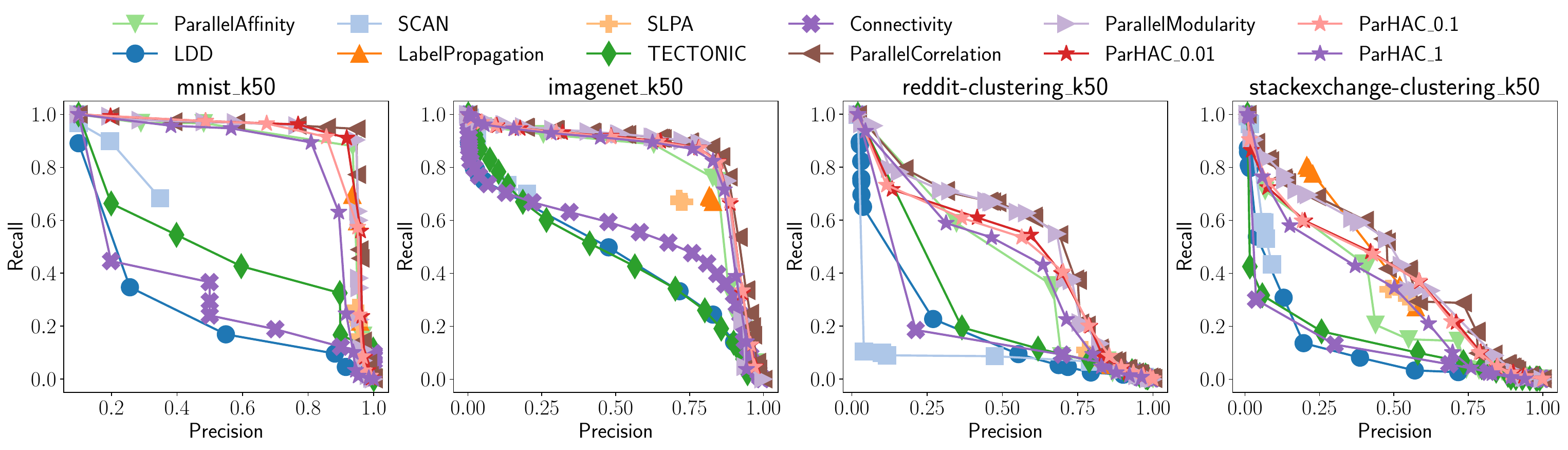}
    \caption{The full Pareto frontier of the precision and recall for the weighted \knn graphs ($k=50$), using \parc{} methods.}
    \label{fig:pr_weighted_full}
\end{figure*}

\section{Compare Modularity Clustering
Implementations}\label{sec:appendix-mod}
\begin{figure*}
    \centering
    \includegraphics[width=\textwidth]{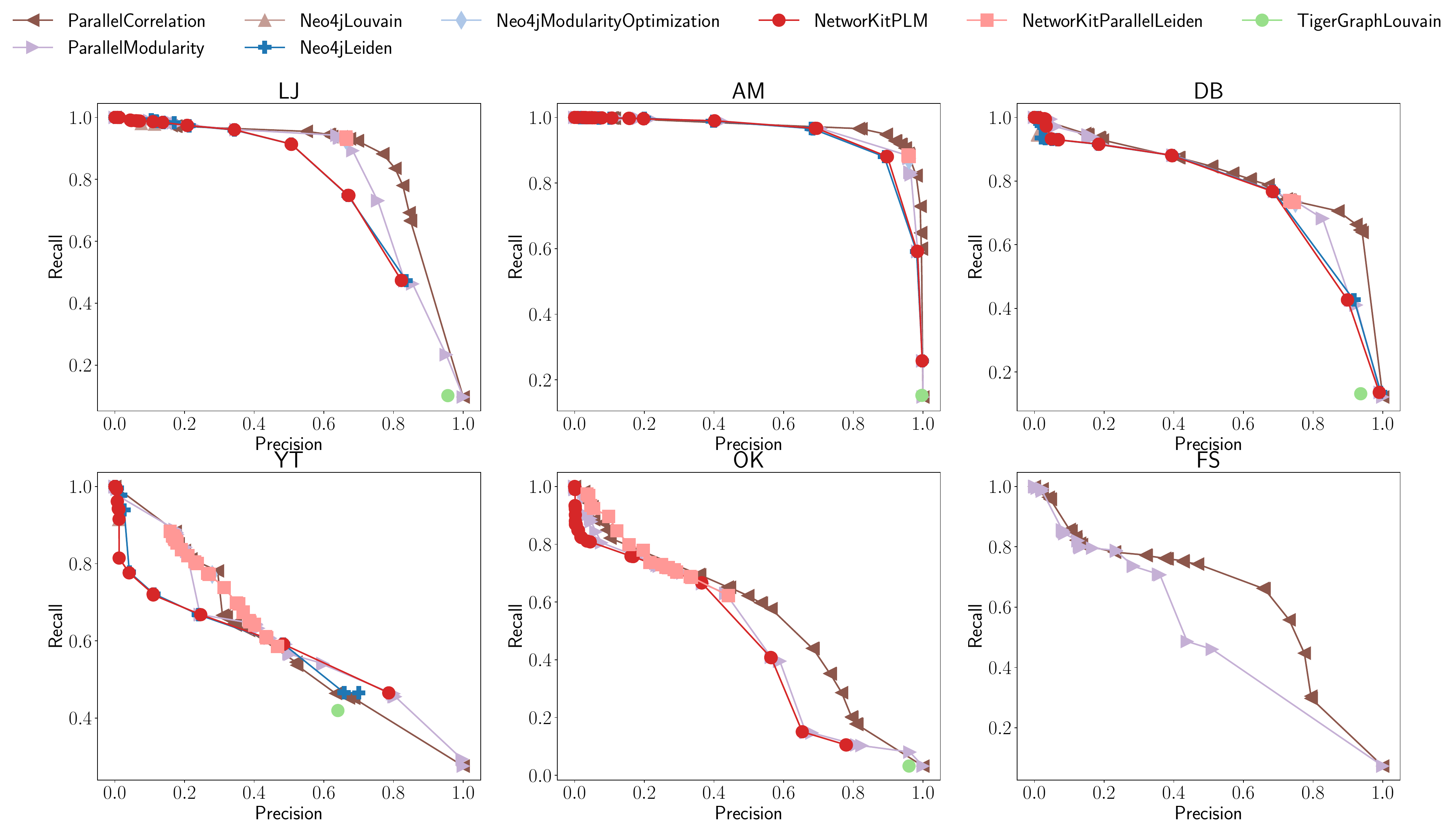}
    \caption{The Pareto frontier of the precision and recall for the unweighted SNAP graphs, using different modularity methods.}
    \label{fig:snap_modularity}
\end{figure*}

\begin{figure*}
    \centering
    \includegraphics[width=\textwidth]{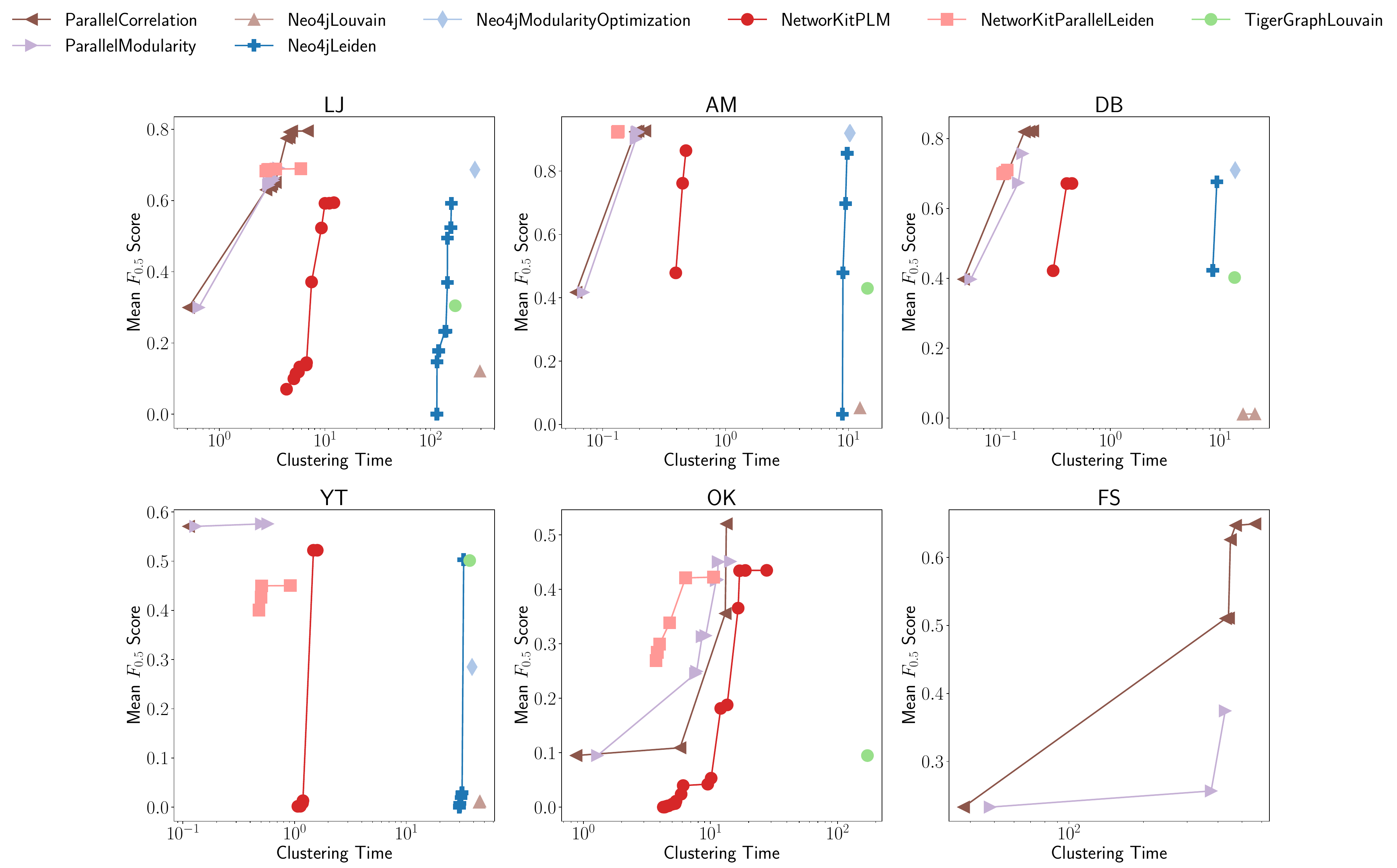}
    \caption{The Pareto frontier of the $F_{0.5}$ and runtime graph for the unweighted SNAP graphs, using different modularity methods.}
    \label{fig:snap_modularity_f1}
\end{figure*}

\begin{figure*}
    \centering
    \includegraphics[width=\textwidth]{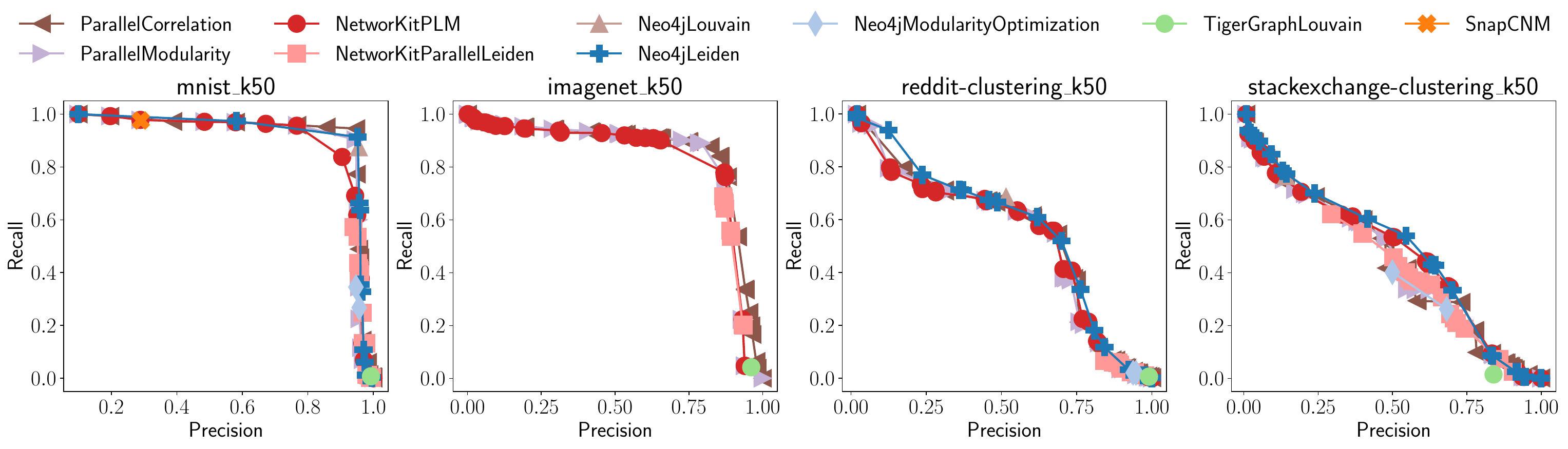}
    \includegraphics[width=\textwidth]{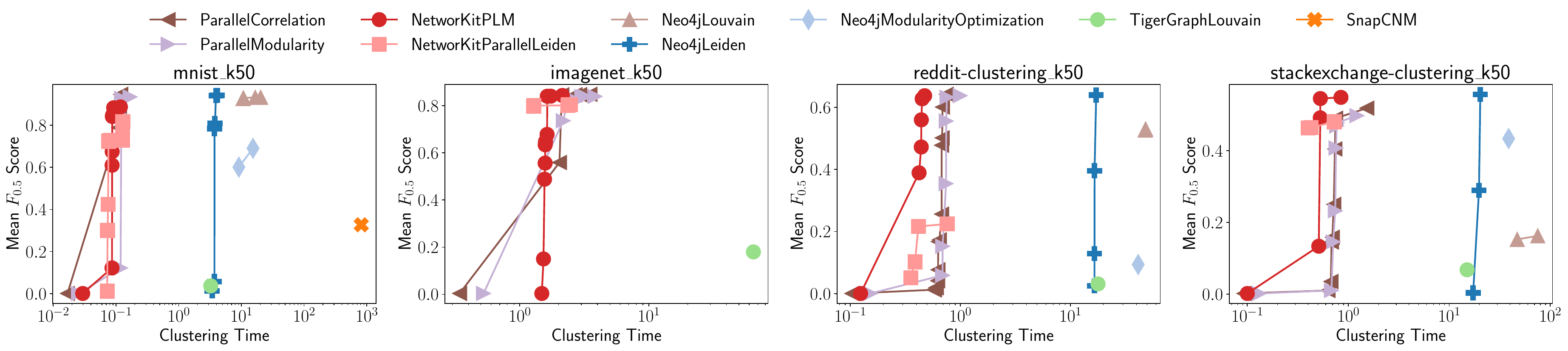}
    \caption{The Pareto frontier graphs for the weighted large \knn graphs ($k=50$), using different modularity methods.}
    \label{fig:large_weight_appendix}
\end{figure*}

\begin{figure*}
    \centering
    \includegraphics[width=\textwidth]{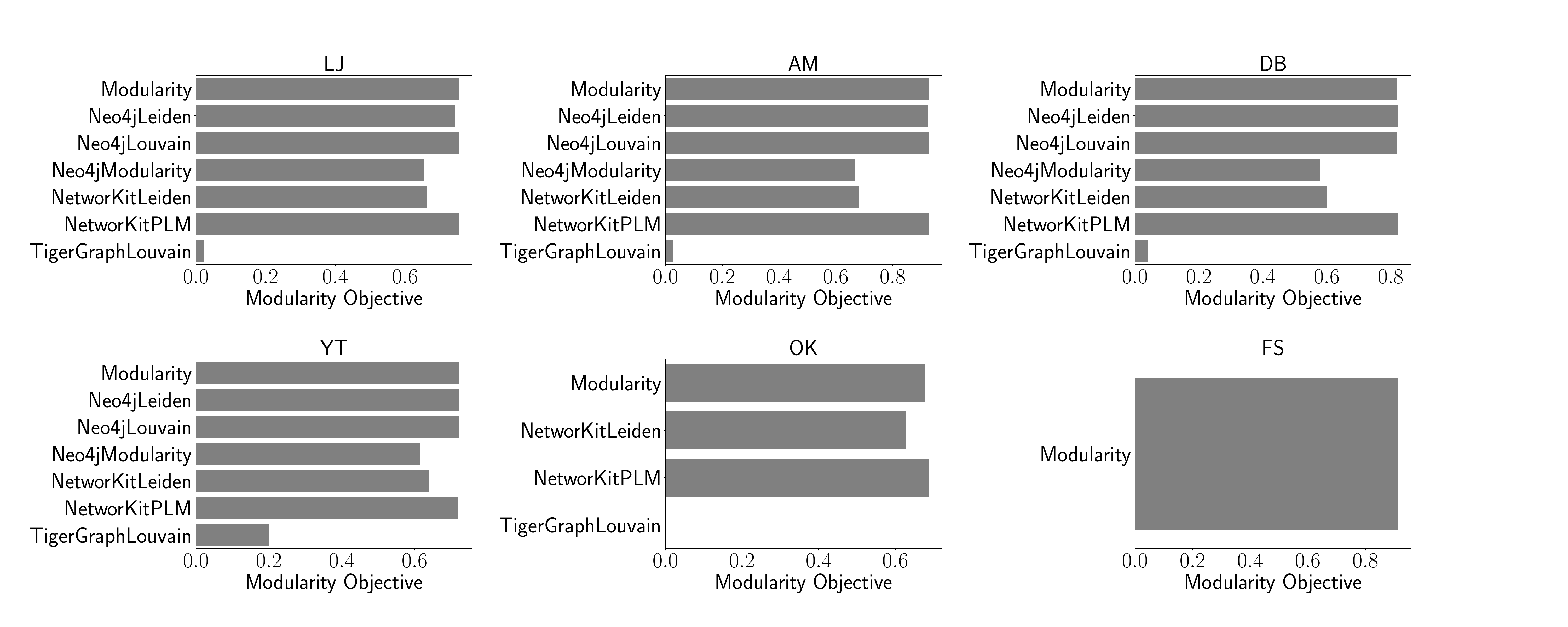}
    \caption{The modularity scores with $\gamma=1$ for the unweighted graphs, using different modularity methods.}
    \label{fig:modularity_scores_all}
\end{figure*}

In \Cref{fig:snap_modularity,fig:snap_modularity_f1,fig:large_weight_appendix}, we present the Pareto frontier plots of the modularity clustering implementations. 
In \Cref{fig:modularity_scores_all}, we show the modularity scores with $\gamma=1$ for the unweighted graphs, using different modularity methods.

\section{Full experiments on SNAP Graphs and Vector Embedding Graphs}~\label{sec:snap_full}
We show the experimental results on all 6 SNAP graphs in \Cref{fig:pr_snap,fig:time_f1_snap}.

\begin{figure*}
    \centering
    \includegraphics[width=\textwidth]{figures/pr_large_weighted_full.pdf}
    \includegraphics[width=\textwidth]{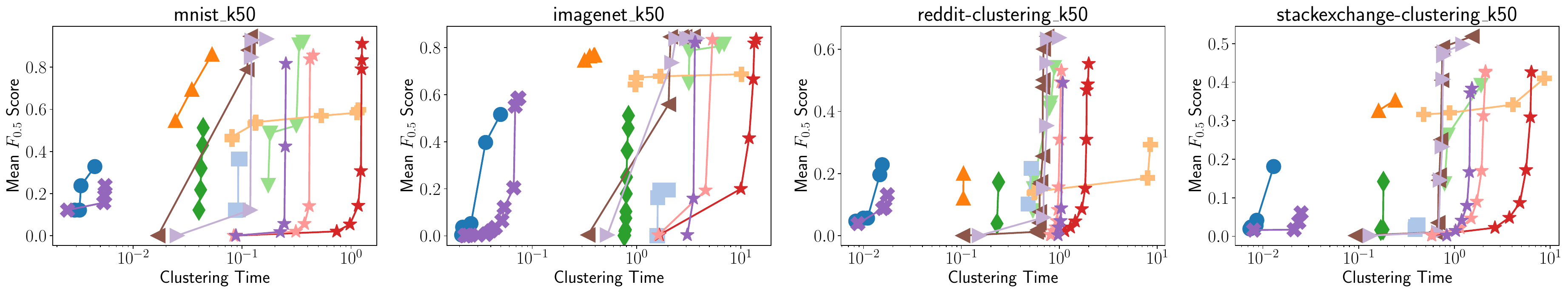}
    \vspace{-2em}
    \caption{\textbf{(Top)} The Pareto frontier of precision and recall for the weighted \knn graphs ($k=50$), using \parc{} methods. \textbf{(Bottom)} The Pareto frontier of $F_{0.5}$ score and clustering time on \knn graphs ($k=50$).
    }
    \label{fig:pr_weighted}
\end{figure*}

\revised{We show the experimental results on all 4 vector embedding graphs in \Cref{fig:pr_weighted}.}

